\documentclass[preprint]{ptephy_v1}
\graphicspath{{./Figs/}}
\usepackage{multirow,epstopdf,url,graphicx,bm}
\usepackage{dcolumn}
\usepackage{here}
\usepackage[normalem]{ulem}

\renewcommand\sout[1]{\bgroup \color{red} \ULdepth=-.5ex \ULset {#1}}
\newcommand{\average}[1]{\ensuremath{\left\langle#1\right\rangle}}

\newcommand{\al}{ a_\eta }
\newcommand{\pf}{ \partial^{\rm F} }

\newcommand{\bx}{ \bm{x} }
\newcommand{\bk}{ \bm{k} }

\newcommand{\fw}{ f_{\rm W} }
\newcommand{\fh}{ f_{\rm H} }
\newcommand{\shw}{ S_{\rm HW} }
\newcommand{\hphi}{ \hat{\phi} }
\newcommand{\hpi}{ \hat{\pi} }

\newcommand{\hca}{ \hat{a}}
\newcommand{\hPhi}{ \hat{\Phi} }
\newcommand{\hPi}{ \hat{\Pi} }

\newcommand{\dd}{ {\rm d} }

\newcommand{\refeq}[1]{(\ref{#1})}
\newcommand{\NTP}{ N_{\rm TP} }
\newcommand{\shws}{ \shw^{\rm sTP} }
\newcommand{\shwp}{ \shw^{\rm pTP} }
\newcommand{\shwa}{ \shw^{\rm aTP} }
\newcommand{\Phik}{\Phi}
\newcommand{\Pik}{\Pi}
\newcommand{\omegak}{\omega}
\preprintnumber{YITP-22-24}
\begin{document}
\title{
Entropy production in longitudinally expanding Yang-Mills field with use of Husimi function --- semiclassical approximation}
\author[1]{Hidefumi Matsuda}
\affil{Department of Physics, Faculty of Science, Kyoto University, Kyoto 606-8502, Japan}
\author[2]{Teiji Kunihiro}
\affil{Yukawa Institute for Theoretical Physics, Kyoto University, Kyoto 606-8502, Japan}
\author[2]{Akira Ohnishi}
\author[3]{Toru T. Takahashi}
\affil{National Institute of Technology, Gunma College, Gunma 371-8530, Japan}
\begin{abstract}
We investigate the possible thermalization process of the highly occupied and weakly coupled 
Yang-Mills fields expanding along the beam axis through an evaluation of the entropy, particle 
number, and pressure anisotropy. 
The time evolution of the system is calculated by solving the equation of motion for the Wigner 
function in the semiclassical approximation with initial conditions mimicking the glasma. 
For the evaluation of the entropy, we adopt the Husimi-Wehrl (HW) entropy, which is obtained by using 
the Husimi function, a positive semidefinite quantum distribution function given by smearing the 
Wigner function. 
By numerical calculations at $g=0.1$ and $0.2$, the entropy production is found to occur together 
with the particle creation in two distinct stages: 
In the first stage, the particle number and the entropy at low longitudinal momenta grow rapidly. 
In the second stage, the particle number and the entropy of higher longitudinal momentum modes 
show slower increase. 
The pressure anisotropy remains in our simulation and implies that the system is still 
out-of-equilibrium. 
\end{abstract}
\maketitle
\section{Introduction}
Experimental studies at RHIC and LHC have provided phenomenological evidences of formation of 
strongly coupled matter soon after the collisions of relativistic heavy ions, and its evolution is 
well described by the hydrodynamics. 
The shear viscosity of the fluid in those successful hydrodynamical models is so 
small~\cite{Perfect2000,Perfect2001_1,Perfect2001_2,Perfect2002} that the entropy produced in the 
hydrodynamical stage is estimated to be about only 10\% of the total 
entropy~\cite{EntropySH2008,EntropyMS2008}. 
Then most of the entropy is expected to be created before the formation of the 
fluid~\cite{EntropyMS2011}.
We have, however, only a poor grasp of the physical mechanism of such early entropy production 
within the underlying Quantum Chromodynamics. 
Thus it is indispensable to elucidate the entropy production mechanism in the pre-hydrodynamic stage 
for a deeper understanding of the outstanding problem of why the hydrodynamics becomes applicable at 
the short time after the collisions, $\tau\sim 0.6-1$ fm/c, known as the early thermalization 
puzzle~\cite{Early2002}.

Shortly after the contact of two nuclei, the produced matter is understood as a highly occupied 
system consisting of weakly coupled but strongly interacting gluons
~\cite{MV1994_1,MV1994_2,MV1994_3,LMW1995_1,LMW1995_2}. 
Such gluonic matter, called a glasma, initially consists of an approximately 
boost invariant color electric and color magnetic fields parallel to the collision axis. 
The fluctuations of this boost invariant fields grow exponentially due to instabilities of the 
Yang-Mills theory
~\cite{Weibel1959,Mrowczynski1988,Mrowczynski1993,Mrowczynski2003,RS2003,RS2004,Arnold2003,
Savvidy1997,Savvidy1978,Iwazaki2008,NO1978,FI2008,FIW2009,InstabilityCYM2006_1,InstabilityCYM2006_2,
InstabilityCYM2006_3,InstabilityCYM2008,InstabilityCYM2009,InstabilityCYM2012FG,
InstabilityCYM2012BSSS,InstabilityCYM2013,Tsutsui2015,Tsutsui2016}. 
The exponential growth of fluctuations should play a role in the isotropization of pressure in the 
glasma~\cite{PressureDEGV2012,PressureEG2013}, and is also expected to drive entropy production in 
the glasma~\cite{Tsukiji2016,Tsukiji2018}. 
Eventually, the grown Yang-Mills field is expected to decay into particles and to form hydrodynamic fluid.

In order to investigate the real-time dynamics of the glasma, a semiclassical method is
widely used~\cite{InstabilityCYM2006_3,InstabilityCYM2012FG,PressureEG2013,Berges2014_1,Berges2014_2,BergesCSS2014,
Tsukiji2018}. 
In this method, the classical field equation of motion is solved starting from the 
initial conditions containing quantum fluctuations, which is implemented to evaluate 
real-time evolution of a Wigner function, a Wigner-Weyl 
transform of a density matrix in terms of the field variables and their conjugate momenta. 
Such semiclassical description can be applied to real-time evolution of highly occupied and weakly 
coupled systems~\cite{BergesCSS2014}, where a quantum effect gives only a small contribution.

It should be noted that some kind of coarse-graining is necessary in order to discuss thermalization 
in terms of the entropy based on the density matrix or the Wigner function or the classical phase 
space distribution function defined microscopically. 
Exact quantum evolution of a density matrix $\hat \rho$ is unitary, and thereby the von-Neumann 
entropy $S=-{\rm Tr}\rho \ln \rho$ stays constant. 
Analogously in classical systems, the phase space distribution function is constant along the 
classical trajectory and the Boltzmann entropy, $S=-\int dxdp/(2\pi)^D f \ln f$ with $D$ and $f$ 
being the number of degrees of freedom and the classical distribution function, stays constant due to 
the Liouville's theorem. 
One of the ways to perform coarse-graining is to use the entanglement entropy~\cite{entangle1,entangle2}, the 
von-Neumann entropy defined by a partially traced density matrix. 
The time evolution of the partially traced density matrix is non-unitary and then the entanglement 
entropy can grow in time. 
However, it is difficult to perform the partial trace for many-body systems such as a field theory. 
Another way is to use the entropy defined by the smeared density matrix or the Wigner function. 
The Husimi function, smeared Wigner function within the allowance of the uncertainty principle, 
is positive semidefinite and can be regarded as a probability density function in the phase space, 
and thus we can define the entropy based on the Husimi function, which we call the Husimi-Wehrl (HW) 
entropy~\cite{Husimi1940,Wehrl1978,Wehrl1979,Kyoto2009}. 
The properties of the HW entropy in- and out-of-equilibrium have been studied analytically in some 
simple models: 
For a harmonic oscillator with a quanta $\hbar \omega$, the HW entropy for the Gibbs ensemble is 
larger than the von-Neumann entropy but tends to 
agree with that in the classical/high-temperature limit ($\hbar/T\to0$)~\cite{Kyoto2009}. 
In an inverted harmonic oscillator, the growth rate of the HW entropy asymptotically converges to the 
Kolmogorov-Sinai (KS) entropy, a sum of positive Lyapunov exponents~\cite{Kyoto2009}, which implies 
that the production of the HW entropy is related to the chaoticity and instabilities in its classical 
counterpart. 
Both results suggest that the HW entropy can be a suitable guide to investigate thermalization at 
least in classical or semiclassical systems.

In this article, we evaluate the evolution of the Wigner function of the highly occupied and weakly 
coupled Yang-Mills fields with initial conditions mimicking the glasma in the semiclassical 
approximation, and analyze its thermalization in terms of the HW entropy that is obtained from the 
evaluated Wigner function. 
The semiclassical approximation method adopted here is essentially the same as the so-called classical 
statistical approximation
~\cite{InstabilityCYM2006_3,InstabilityCYM2012FG,PressureEG2013,Berges2014_1,Berges2014_2,BergesCSS2014,Tsukiji2018} 
in the sense that the initial conditions are sampled by Monte Carlo method 
and the classical equations of motion are solved.
Our simulation is performed in the $\tau-\eta$ coordinate system, which represents a system expanding 
along the beam axis at the speed of light. 
In Refs.~\cite{Tsukiji2016,Tsukiji2018}, the authors numerically showed that the HW entropy does grow 
definitely by the classical dynamics of the Yang-Mills field, and that the growth rate of the HW 
entropy is related to the intrinsic dynamics of the Yang-Mills theory, such as the chaoticity and 
instabilities. 
It should be noted, however, that the previous studies were only made in a static geometry with a 
focus on the thermalization by the intrinsic dynamics of the Yang-Mills field. 
Thus it would be intriguing to examine whether their findings are robust enough that they persist 
in an expanding geometry. 
The present work is an extension of the previous analysis to the expanding system with some technical 
improvements:
We give an improved definition of the HW entropy in field theories that resolves two problems left 
unsolved in Refs.~\cite{Tsukiji2016,Tsukiji2018}, over-counting gauge degrees of freedom and an 
ambiguity in choice of smearing parameters. 
Another important development in this article is the evaluation of the particle number and its 
relevance to the entropy production. 
The initial condition given as the glasma-like one with quantum fluctuations may be described as 
the coherent state. 
Then the deviation from the coherent state is realized in the subsequent time evolution. 
We give an operator representation of the particle number created due to this deviation, and 
investigate the relation of the entropy and particle number production. 
Moreover we provide a theoretical basis of the numerical method for a precise evaluation of the HW 
entropy by the test particle method, which was proposed in Refs.~\cite{Tsukiji2016}.

This article is organized as follows. 
In Sec.~\ref{Sec:Theory}, we introduce the semiclassical description of the real-time evolution of 
quantum systems based on the Wigner function, and give the definition of the HW entropy, using a 
simple example: one-dimensional quantum mechanics. 
In Sec.~\ref{Sec:Scalar}, we show how to numerically evaluate the HW entropy of a semiclassical 
field, using the scalar theory in the Minkowski spacetime. 
In Sec.~\ref{Sec:YM}, we investigate the dynamical production of the HW entropy in the semiclassical 
evolution of the $SU(2)$ Yang-Mills field in the expanding geometry. 
We also study other observables, pressure and particle number, and discuss the relation between the 
HW entropy and them. 
In Sec.~\ref{Sec:Summary}, we summarize this article.

\section{Formalism}\label{Sec:Theory}
In this section, we introduce the semiclassical description of the real-time evolution of quantum 
systems based on the Wigner function, and give the definition of the HW entropy, using a 
one-dimensional harmonic oscillator whose  Hamiltonian reads 
$H=\hat{p}^2/2+\omega^2\hat{x}^2/2=\omega(\hat{a}^{\dag}\hat{a}+1/2)$ with 
$\hca=(\omega\hat{x}+i\hat{p})/\sqrt{2\omega}$.

\subsection{Semiclassical description of Wigner function}\label{Sec:CSS}
The Wigner function is defined as the Wigner-Weyl transform of the density matrix $\hat{\rho}(t)$,
\begin{align}
f_{\rm W}(x,p) \equiv \int {\rm d}y \average{x+\frac{y}{2} \mid \hat{\rho}(t) \mid x-\frac{y}{2}}e^{-ipy}\ .
\end{align}
To describe the semiclassical evolution of the Wigner function, we use the classical limit of the 
von-Neumann equation, 
\begin{align}
\frac{\partial}{\partial t}f_{\rm W}(x,p)=\lim_{\hbar\to0}\{\{ H, \fw(x,p) \}\}\ ,\label{Eq:CvN}
\end{align}
where $\{\{,\}\}$ denotes the Moyal bracket. 
The Moyal bracket can be written as a power series of $\hbar^2$,
\begin{align}
\{\{ H, \fw(x,p) \}\}=\{ H, \fw(x,p) \}+\mathcal{O}(\hbar^2)\ ,\label{Eq:CvN2}
\end{align}
where $\{,\}$ denotes the Poisson bracket. 
In the classical limit ($\hbar\rightarrow 0$), therefore, Eq.~\refeq{Eq:CvN} 
reads the same form as the Liouville equation that is an evolution equation of a classical 
distribution function,
\begin{align}
\frac{\partial}{\partial t}f_{\rm W}(x,p)=\{ H, \fw(x,p) \}\ .\label{Eq:Liouville}
\end{align}
By combining the classical evolution equation given by Eq.~\refeq{Eq:Liouville} with the 
semiclassical initial condition given later in Eq.~\refeq{Eq:Coherent}, we can describe the 
time-dependent Wigner function within the semiclassical approximation, where quantum fluctuation 
effects up to $\mathcal{O}(\hbar^1)$ are included.

We calculate an expectation value of the given observable $\hat{\mathcal{O}}(\hat{x},\hat{p})$ 
through the following relation,
\begin{align}
\average{\hat{\mathcal{O}}(\hat{x},\hat{p})}=
{\rm Tr}\left(\hat{\mathcal{O}}(\hat{x},\hat{p})\hat{\rho}\right)
=\int d\Gamma\ \mathcal{O}_{\rm W}(x,p) f_{\rm W}(x,p)\ ,\label{Eq:CSSob}
\end{align}
where $d\Gamma=dxdp/(2\pi)$ is the integration measure and $\mathcal{O}_{\rm W}(x,p)$ is the 
Wigner-Weyl transform of $\hat{\mathcal{O}}(\hat{x},\hat{p})$,
\begin{align}
\mathcal{O}_{\rm W}(x,p)=\int {\rm d}y \average{x+\frac{y}{2}| \hat{\mathcal{O}}(\hat{x},\hat{p}) |x-\frac{y}{2}}e^{-ipy}\ .
\end{align}
In actual calculations in field theories, we may need to subtract the vacuum expectation value and 
others to obtain the well-defined observables as discussed later.

\subsection{Wigner function of a coherent state}\label{Sec:CSS}
We use a coherent state for the initial condition in actual calculations discussed later, since a 
description with use of a coherent state is useful when the semiclassical approximation is valid.  
A coherent state is an eigenstate of the annihilation operator 
$\hca|\alpha\rangle=\alpha|\alpha\rangle$, and is represented as 
$|\alpha\rangle=e^{-|\alpha|^2/2}e^{\alpha \hca^\dagger}|0\rangle$ with 
$|0\rangle$ being the perturbative vacuum state; 
$\hca\vert 0\rangle=0$. 
The coherent state satisfies the minimum uncertainty relation, $\Delta x\Delta p=\frac{1}{2}$.
\
With use of the density matrix $\hat{\rho}:=\vert \alpha\rangle\langle \alpha\vert$, the Wigner 
function of the coherent state $|\alpha\rangle$ is given by
\begin{align}
f_{\rm W}(x,p)=&\int {\rm d}y
\average{x+\frac{y}{2} \mid \alpha}
\average{\alpha \mid x-\frac{y}{2}}e^{-ipy}
=2\exp\left[
-\frac{(x-\bar{x})^2}{2(\Delta x)^2}
-\frac{(p-\bar{p})^2}{2(\Delta p)^2}
\right]
\ ,\label{Eq:Coherent}
\end{align}
where 
$\bar{x}$ and $\bar{p}$ are the expectation values of the position and momentum, respectively, and 
are related to the eigenvalue as $\alpha=(\omega\bar{x}+i\bar{p})/\sqrt{2\omega}$.
The variances are given by 
$(\Delta x)^2:=\langle \alpha \vert (\hat{x}-\bar{x})^2\vert \alpha\rangle=1/(2\omega)$ and 
$(\Delta p)^2:=\langle \alpha \vert (\hat{p}-\bar{p})^2\vert \alpha\rangle=\omega/2$.


\subsection{Husimi-Wehrl entropy}\label{Sec:HW}
The Husimi function is obtained by smearing the Wigner function in the phase space within the 
allowance of the uncertainty principle, 
\begin{align}
\fh(x,p,\sigma) \equiv \int{\rm d} \Gamma' f_{\rm W}(x',p')G(x-x',p-p',\sigma)\ ,
\end{align}
where  $G(x,p,\sigma)$ is the Gaussian smearing function,
\begin{align}
G(x,p,\sigma)=2 e^{-\sigma x^2-p^2/\sigma}\ ,\label{Eq:smearing}
\end{align}
where $\sigma$ is the smearing parameter. 
The Husimi function is normalized in the phase space as $\int d\Gamma \fh(x,p,\sigma)=1$. 
In addition, it is given as the expectation value of the density matrix in a coherent state 
$|\alpha;\sigma\rangle$ defined as the eigenstate of the annihilation operator 
$\hca=(\sigma\hat{x}+i\hat{p})/\sqrt{2\sigma}$, 
$\fh(x,p,\sigma)=\langle\alpha;\sigma\!\mid\!\hat{\rho}\!\mid\!\alpha;\sigma\rangle$, and is positive 
semidefinite unlike the Wigner function. 
Therefore, the Husimi function can be regarded as a probability distribution function in the phase
space. 
Then, we can define the Husimi-Wehrl entropy as
\begin{align}
\shw(\sigma)\equiv-\int {\rm d}\Gamma \fh(x,p,\sigma)\ln{\fh(x,p,\sigma)}\ .
\end{align}
While there exists an ambiguity of the choice of the smearing parameter $\sigma$ in the Husimi 
function, we unambiguously fix $\sigma$ by imposing physically natural requirement in the later 
section.

\section{Husimi-Wehrl Entropy of scalar field in non-expanding geometry}\label{Sec:Scalar}
Before proceeding with the study of the Yang-Mills field in the expanding geometry, we show how to numerically 
evaluate the HW entropy of a semiclassical field, using the scalar theory 
in Minkowski space-time as an example. 
All quantities in this section are normalized by a spatial lattice spacing $a$.

\subsection{Scalar Field Theory on Lattice}\label{Sec:ScalarTheory}
We consider the massless $\phi^4$ theory on a $L^3$ lattice whose Hamiltonian is given by
\begin{align} 
H=\sum_{\bx} 
\left[ \frac{1}{2}\hpi(x) ^2 - \frac{1}{2}\left( \bm{\nabla}^{\rm F}\hphi(x) \right)^2 + \frac{\lambda}{4!}\hphi^4(x) \right]\ ,\label{Eq:H}
\end{align} 
where $\bm{\nabla}^{\rm F}=(\partial^{\rm F}_1,\partial^{\rm F}_2,\partial^{\rm F}_3)$ denotes a 
forward difference operator, and $\hpi=\dot{\hphi}$ is the canonical conjugate variable of $\hphi$. 
Then, field variables for the free scalar field ($\lambda=0$) on the lattice after the second 
quantization are given by
\begin{align}
\hat{\phi}(x)&= \frac{1}{\sqrt{L^3}}\sum_{\bk}\hat{\phi}_{\bk}e^{i\bk\cdot\bx}
              = \frac{1}{\sqrt{L^3}}\sum_{\bk}\frac{1}{\sqrt{2\omega_{\bk}}} 
\left(  \hca_{\bk}e^{-i\omega_{\bk} t} + \hca^\dagger_{-\bk}e^{i\omega_{\bk} t} \right)e^{i\bk\cdot\bx}\ ,\label{Eq:phi}\\
\hat{\pi}(x) &= \frac{1}{\sqrt{L^3}}\sum_{\bk}\hat{\pi}_{\bk}e^{i\bk\cdot\bx}
              = \frac{-i}{\sqrt{L^3}}\sum_{\bk}\sqrt{\frac{\omega_{\bk}}{2}} 
\left(  \hca_{\bk}e^{-i\omega_{\bk} t} - \hca^\dagger_{-\bk}e^{i\omega_{\bk} t} \right)e^{i\bk\cdot\bx}\ ,\label{Eq:pi} 
\end{align}
where $(\hat{\phi}_{\bk}, \hat{\pi}_{\bk})$ are the Fourier transform of 
$(\hat{\phi}(x), \hat{\pi}(x))$, $\hca_{\bk}$ $(\hca^\dagger_{\bk})$ is the annihilation (creation) 
operator for the free scalar field satisfying 
$\left[\hca_{\bk}, \hca^{\dagger}_{\bk'}\right] = \delta_{\bk,\bk'}$, and 
$\omega_{\bk}=2\sqrt{\sin^2(k_1/2)+\sin^2(k_2/2)+\sin^2(k_3/2)}$ is the eigenfrequency of the free 
field mode with momentum $\bk$ on the lattice.

The free part of the scalar theory can be interpreted as a set of harmonic oscillators, whose 
Hamiltonian is given by
\begin{align}
H_{\rm free}
=\sum_{\bk}\,\omega_{\bk}\left( \hca^\dagger_{\bk}\hca_{\bk} + \frac{1}{2} \right)
=\sum_{\bk} H^{\rm h.o.}_{\bk}
\,,\ 
H^{\rm h.o.}_{\bk}
=\frac{1}{2}\hPi_{\bk}^2+\frac{1}{2}\omega^2_{\bk}\hPhi_{\bk}^2
\,,\label{Eq:Hfree_new}
\end{align}
where we have introduced the new canonical variables for each momentum mode, 
$(\hPhi_{\bk},\hPi_{\bk})$, utilizing the annihilation and creation operators as
\begin{align}
\hPhi_{\bk} \equiv  \frac{1}{\sqrt{2\omega_{\bk}}}  \left[ \hca_{\bk}+\hca^{\dagger}_{\bk}\right]\ ,\ \ \
\hPi _{\bk} \equiv  -i\sqrt{\frac{\omega_{\bk}}{2}} \left[ \hca_{\bk}-\hca^{\dagger}_{\bk}\right]\ ,
\end{align}
or accordingly,
\begin{align}
\hca_{\bk}=(\omega_{\bk}\hPhi_{\bk}+i\hPi_{\bk})/\sqrt{2\omega_{\bk}}. 
\label{eq:field-annih-cano-rel}
\end{align}

Now we can consider the Winger function $\fw(\{\Phik,\Pik\})$ expressed in terms of 
$\Phi_{\bk}$ and $\Pi_{\bk}$. 
As the initial condition, we adopt the coherent state $|\{\alpha_{\bk}\}\rangle$ that is the 
eigenstate of annihilation operators for the free scalar field, $\{\hat{a}_{\bk}\}$, and is given by 
the product of coherent states of each mode, 
$|\{\alpha_{\bk}\}\rangle=\prod_{\bk}|\alpha_{\bk}\rangle$. 
Accordingly, the density matrix is expressed as 
$\hat{\rho}(t)=|\{\alpha_{\bk}\}\rangle\langle \{\alpha_{\bk}\}|
=\prod_{\bk}|\alpha_{\bk}\rangle\prod_{\bk'}\langle\alpha_{\bk'}\vert$. 
Then, as shown in Eq~\refeq{Eq:Coherent}, the Wigner function at the initial time is given by the 
Gaussian function,
\begin{align}
\fw(\{\Phik,\Pik\})
&=\prod_{\bk} \int {\rm d}\Phi'_{\bk}
  \average{\Phi_{\bk}+\frac{\Phi'_{\bk}}{2}\,\Big|\,\alpha_{\bk}}
  \average{\alpha_{\bk}\,\Big|\,\Phi_{\bk}-\frac{\Phi'_{\bk}}{2}}e^{-i\Pi_{\bk}\Phi'_{\bk}}\nonumber\\
&=\prod_{\bk} 2
\exp\left[
-\frac{(\Phi_{\bk}-\bar{\Phi}_{\bk})^2}{2(\Delta \Phi_{\bk})^2}
-\frac{(\Pi _{\bk}-\bar{\Pi }_{\bk})^2}{2(\Delta \Pi_{\bk})^2}
\right]\,,
\label{Eq:Wigini}
\end{align}
where the expectation values, $\bar{\Phi}_{\bk}$ and $\bar{\Pi}_{\bk}$, 
are the real and imaginary parts of the coherent state eigenvalue, 
$\alpha_{\bk}=(\omega_{\bk}\bar{\Phi}_{\bk}+i\bar{\Pi}_{\bk})/\sqrt{2\omega_{\bk}}$, 
and the variances, $(\Delta \Phi_{\bk})^2$ and $(\Delta \Pi_{\bk})^2$, are determined by the 
eigenfrequency $\omega_{\bk}$ as $\omega_{\bk}=1/[2(\Delta \Phi_{\bk})^2]=2(\Delta \Pi_{\bk})^2$.
It is noted that the spread of the initial Wigner 
function characterized by the widths, 
$\Delta \Phi_{\bm{k}}$ and $\Delta \Pi_{\bm{k}}$, 
corresponds to the zero-point oscillation of the 
vacuum if fluctuations of fields are regarded as 
particles.

We further define a Husimi function $f_H$ as a smeared Wigner function with the smearing parameters 
$\{\sigma_{\bk}\}$ set to eigenfrequencies $\{ \omega_{\bk} \}$, 
\begin{align}
\fh(\{\Phik,\Pik,\omegak\})=
\int \mathcal{D}\Gamma' \fw(\{\Phik',\Pik'\})
 G(\{\Phik-\Phik',\Pik-\Pik',\omegak\})\ ,\label{Eq:Husimi}
\end{align}
where $\mathcal{D}\Gamma=\prod_{\bk}d\Phi_{\bk}d\Pi_{\bk}/(2\pi)$ is an integration measure 
and $G(\{\Phik,\Pik,\omegak\})$ is the smearing function defined as the product of single Gaussian 
smearing functions given in Eq.\refeq{Eq:smearing}, 
\begin{align}
G(\{\Phik,\Pik,\omegak\}) =  \prod_{\bk} G(\Phi_{\bk},\Pi_{\bk},\omega_{\bk})\ .
\end{align}
The HW entropy is finally given by 
\begin{align}
\shw(\{\omegak\})=-\int \mathcal{D}\Gamma \fh(\{\Phik,\Pik,\omegak\})
         \ln{\fh(\{\Phik,\Pik,\omegak\})}\ .\label{Eq:HWs}
\end{align}
We here comment on the smearing parameter choice shown above.
The HW entropy $\shw(\{\omegak\})$ defined with the smearing parameter $\{\omegak\}$ exhibits two 
remarkable and physically natural features, as shown in App.~\ref{App:ChoiceOfSigma}, at 
$\lambda=0$: 
The HW entropy $\shw(\{\omegak\})$ agrees with the von-Neumann entropy 
$S_{\rm vN}=-{\rm Tr}(\rho\ln \rho)$ in the high-temperature limit, and the HW entropy $\shw(\{\omegak\})$ per 
degrees of freedom in vacuum is unity, the minimum of the HW entropy. 
These two features may justify to adopt the definition in weak coupling calculations. 
In the later discussion of this section, we omit $\{\omega\}$ in the expression of the HW 
entropy of the scalar field: $\shw$. 

\subsection{Numerical method}
\subsubsection{Evaluation of Wigner function using test particle method}
\ \\

In the actual evaluation of the Wigner function, we use the test particle (TP) method in which the 
Wigner function is approximated by a sum of the delta functions,
\begin{align}
\fw^{\rm TP}(\{\Phik,\Pik\})
=\frac{1}{\NTP}\sum_i \prod_{\bk} (2\pi)\delta(\Phi_{\bk}-\Phi_{\bk,i}(t))\delta(\Pi_{\bk}-\Pi_{\bk,i}(t))\ ,\label{Eq:WignerTP}
\end{align}
where each delta function specifies the coordinate of an independent particle $(\Phi_{\bk,i},\Pi_{\bk,i})$ which is generated so 
as to sample the classical field configurations (test particle configurations) according to the 
distribution of the Wigner function. 
The test particle method assumes the positive semi-definiteness of the Wigner function, which is 
certainly true in the time evolution of the Wigner function according to the Liouville equation 
shown in Eq.~\refeq{Eq:Liouville} starting from the positive definite initial condition like 
Eq.~\refeq{Eq:Wigini}. 
The Winger function given in Eq.~\refeq{Eq:WignerTP} should give an accurate sampling of the original 
Wigner function in the large $\NTP$ limit. 

The test particles at initial time are generated according to the initial Wigner function given in 
Eq~\refeq{Eq:Wigini},
\begin{align}
\Phi_{\bk,i} =\bar\Phi_{\bk}+\sqrt{\frac{1}{2\omega_{\bk}}} \xi_{\bk,i}^R\,,
\ 
\Pi_{\bk,i}  =\bar\Pi_{\bk} +\sqrt{\frac{\omega_{\bk}}{2}}   \xi_{\bk,i}^I\,,
\label{Eq:ini}
\end{align}
where $\xi_{\bk,i}^{R,I}$ are the random numbers obeying the normal Gaussian distribution. 
The time evolution of each test particle is obtained by the classical equation of motion, 
$\dot{\phi}(x)=\partial H/\partial \pi(x), \dot{\pi}(x)=-\partial H/\partial \phi(x)$.
In the later discussion, we calculate an expectation value of 
a given observable $\hat{\mathcal{O}}(\{\hat{\Phi},\hat{\Pi}\})$, 
such as pressure and particle number, 
by using the test particle expression
of the Wigner function given in Eq.~\eqref{Eq:CSSob},
\begin{eqnarray}
\average{\hat{\mathcal{O}}(\{\hat{\Phi},\hat{\Pi}\})}=
\frac{1}{N_{\rm TP}}\sum_i \mathcal{O}
(\{\Phi_i,\Pi_i\})\ .
\end{eqnarray}

\subsubsection{Husimi-Wehrl entropy in test particle method}
\ \\

By substituting the Wigner function given in Eq.~\refeq{Eq:WignerTP} into
Eq.~\refeq{Eq:Husimi}, we obtain the Husimi function in the test particle method, 
\begin{align}
&\fh^{\rm TP}(\{\Phik,\Pik\})=\frac{1}{\NTP}\sum_i 
\prod_{\bk}G_{\bk,i}(\Phi_{\bk},\Pi_{\bk})\,,
\label{Eq:fHTP}\\
&G_{\bk,i}(\Phi_{\bk},\Pi_{\bk})
=G(\Phi_{\bk}-\Phi_{\bk,i}(t),\Pi_{\bk}-\Pi_{\bk,i}(t),\omega_{\bk})\ .
\end{align}
Then the HW entropy in the test particle method is given by
\begin{align}
\shw^{\rm TP}=&-\int D\Gamma\left[\frac{1}{\NTP}\sum_i \prod_{\bk }
G_{\bk,i}(\Phi_{\bk},\Pi_{\bk})
\right]
\ln{\left[\frac{1}{\NTP}\sum_j \prod_{\bk'}
G_{\bk',j}(\Phi_{\bk'},\Pi_{\bk'})\right]}
\,.\label{Eq:Stp}
\end{align}

The HW entropy is evaluated in the following three steps in our test particle method which 
is elaborated from that adopted in the previous work. 
In the first step, we apply identical test particle sets to functions both inside and outside the 
logarithmic function in Eq.~\refeq{Eq:Stp}, and evaluate the HW entropy.
We call this prescription the "single test particle method (sTP)", and denote the HW entropy obtained 
by the sTP method as $\shws$. 
In the second step, we compute the HW entropy using two different test particle sets for functions 
inside and outside the logarithmic function in Eq.~\refeq{Eq:Stp}. 
We call the second prescription the "parallel test particle method (pTP)", 
and denote the HW entropy obtained by the pTP method as $\shwp$.
Finally, we estimate the HW entropy by averaging $\shws$ and $\shwp$, $\shwa \equiv (\shws+\shwp)/2$.
As proven in App.~\ref{App:ne_tpm}, 
we obtain the following inequality,
\begin{align}
\shws < \shw < \shwp\ .\label{Eq:sptp}
\end{align}
by assuming that numerical 
errors of the Husimi function, 
$\Delta^{\rm TP} = \fh^{\rm TP}-\fh \propto 1/\sqrt{\NTP}$, 
is sufficiently small due to the large value of $\NTP$,
and that the odd-order contributions of 
$\Delta^{\rm TP}$ to $\shws$ and $\shwp$ disappear due
to the numerical error cancellation,
Moreover, it is also shown in App.~\ref{App:ne_tpm} that the numerical errors proportional to 
$\NTP^{-1}$ cancel out each other in $\shwa$, and $\shwa$ only contains $\mathcal{O}(\NTP^{-2})$ 
errors.
Thus, for the sufficiently large $\NTP$, the following nice equality holds,
\begin{align}
\shwa = \shw + \mathcal{O}(\NTP^{-2})\ .\label{Eq:atp}
\end{align}
In the next section, we numerically calculate $\shws$, $\shwp$ and $\shwa$ at different numbers of 
test particles, and discuss the validity of Eq.~\refeq{Eq:sptp} and Eq.~\refeq{Eq:atp}.

\subsubsection{Product ansatz}
\ \\

We need to make a further approximation, assuming the product ansatz for the Wigner function in order 
to obtain the HW entropy in Eq.~\refeq{Eq:Stp} in the test particle method.
While the Husimi function is equivalent to the expectation value of the density matrix in the 
coherent state and takes a value $0 \leq \fh \leq 1$, the Husimi function in the test particle 
method, Eq.~\refeq{Eq:fHTP}, takes a value $\fh \geq 2^{N_D}/\NTP$ at one of the test particle 
phase-space coordinates, $(\Phi_{\bk},\Pi_{\bk})=(\Phi_{\bk,i},\Pi_{\bk,i})$, with $N_D=L^3$ being 
the number of degrees of freedom. 
Then the required number of test particles is $\NTP > 2^{N_D}$ in order to respect the $\fh$ range.
As in this example, in order to cover the $2N_D$ dimensional phase space, we need a huge number of 
test particles.
We cannot prepare such sufficiently large numbers of test particle configurations for 
large $L$. 

To circumvent this practical problem, we assume the product ansatz for the Wigner function, 
\textit{a la} Hartree-Fock approach,
\begin{align}
\fw(\{\Phik,\Pik\}) = \prod_{\bk} \fw^{\bk}(\Phi_{\bk},\Pi_{\bk})\ ,\ \ \ 
\fw^{\bk}(\Phi_{\bk},\Pi_{\bk})
=\frac{1}{\NTP}\sum_i (2\pi)\delta(\Phi_{\bk}-\Phi_{\bk,i})\delta(\Pi_{\bk}-\Pi_{\bk,i})\ ,
\end{align}
which means that there is no correlation between the wave functions of 
different momentum modes. 
Under the product ansatz, 
the Husimi function is also nicely expressed 
as the product of that of the single momentum mode,
\begin{align}
\fh(\{\Phik,\Pik\})=\prod_{\bk} \fh^{\bk}(\Phi_{\bk},\Pi_{\bk}),\ \ \ 
\fh^{\bk}(\Phi_{\bk},\Pi_{\bk})
=\frac{1}{\NTP}\sum_i G_{\bk,i}(\Phi_{\bk},\Pi_{\bk})\,,
\label{Eq:ProductAnsatzH}
\end{align}
which allows us to treat the multiple integration in Eq.~\refeq{Eq:Stp} as the sum of double 
integrations,
\begin{align}
\shw=&\sum_{\bk} \shw^{\bk}
\,,\ 
\shw^{\bk}=-\int \frac{d\Phi_{\bk} d\Pi_{\bk}}{(2\pi)^3} 
\fh^{\bk}(\Phi_{\bk},\Pi_{\bk})
\ln \fh^{\bk}(\Phi_{\bk},\Pi_{\bk})
\label{Eq:shwpa}
\end{align}

The product ansatz does not necessarily hold in interacting systems,
since the interaction generates correlations between different modes.
The product ansatz for $\fh$ in Eq.~\eqref{Eq:ProductAnsatzH}
tells us that $\fh^{\bk}$ is a partially traced distribution function,
\begin{align}
\fh^{\bk}(\Phi_{\bk},\Pi_{\bk})
=\int \prod_{\bk'\not=\bk}\frac{d\Phi_{\bk'}d\Pi_{\bk'}}{2\pi} \fh(\{\Phik,\Pik\})\,.
\end{align}
Thus while we aim at calculating the entropy from the coarse-graining, the HW entropy with the 
product ansatz given in Eq.~\eqref{Eq:shwpa} is found to overestimate it by the amount of the 
entropy resulting from the loss of the correlation caused by the partial trace of $\fh$.
So far, the entropy increase caused by the use of the product ansatz
have been tested only for a few dimensional quantum system~\cite{Tsukiji2016},
and the entropy increase in such a case is found to be around 20\%
of that from only the coarse-graining.
However, the product ansatz is a kind of a mean-field approximation 
like the Hartree(-Fock) approximation, and is expected to provide a 
good approximation for a system with large degrees of freedom, such as a highly 
occupied system, where fluctuations should be small.

\subsection{Numerical results}\label{Sec:NumScalar}
\begin{figure}[H]
\centering\includegraphics[width=80mm,bb=0 0 720 504]{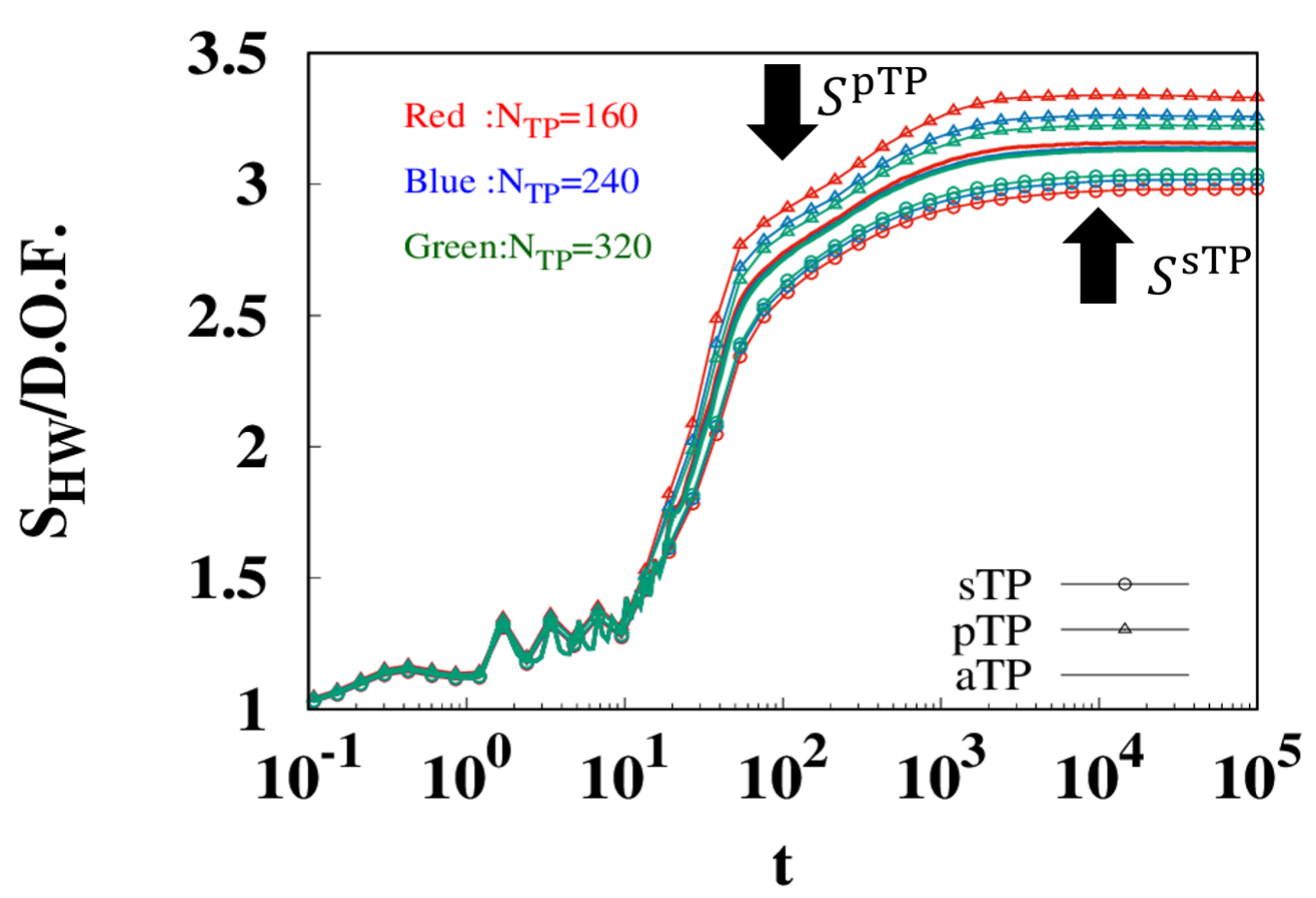}\\
\caption{The time evolution of the HW entropy of the scalar field 
with $\lambda=1$ on the $16^3$ lattice.
The red, blue and green line show the results 
for $\NTP=160$, $240$ and $320$, respectively. 
The lines with circles, triangles, and no symbols
 show the HW entropy evaluated by the sTP method ($\shws$),
the pTP method ($\shwp$), and their average ($\shwa$),
respectively. 
}
\label{Fig:entropyS}
\end{figure}
Here we show the numerical results of the time evolution of the HW entropy  
of the scalar field theory using the product ansatz
with the coupling constant $\lambda=1$
on the $16^3$ lattice with the periodic boundary condition.
The initial positions of test particles in the phase space
are generated according to the initial Wigner function given in Eq~\refeq{Eq:Wigini}, 
where the initial macroscopic fields $(\bar\Phi_{\bk},\bar\Pi_{\bk})$ are given so as to satisfy the following simple initial condition, 
\begin{align}
\bar{\Phi}({\bx})=5\sin{\left(\frac{\pi}{4}(x+y)\right)}\ ,\ \ \ \bar{\Pi}(\bx)=0\ .
\end{align}
The classical equation of motion for each test particle is solved by leapfrog integration. 
The number of test particles is taken as $\NTP=160$, $240$ and $320$.

Figure~\ref{Fig:entropyS} shows the evolution of
$\shws$, $\shwp$ and $\shwa$ per degrees of freedom at $\NTP=160$, $240$ and $320$.
It is found that $\shws<\shwp$ always holds and the difference
between $\shws$ and $\shwp$ becomes smaller as $\NTP$ increases, 
which is consistent with the inequality shown in Eq.~\refeq{Eq:sptp}. 
Moreover, the $\NTP$ dependence of $\shwa$ at $\NTP=160$, $240$ and $320$
is much smaller than those for $\shws$ and $\shwp$, and is only less than 2\%, 
which indicates that the equality shown in Eq.~\refeq{Eq:atp} holds.
In what follows, we adopt $\shwa$ as the estimation of $\shw$, $\shw=\shwa$.

The time evolution of the HW entropy shows the relaxation processes from a coherent state to thermal 
equilibrium. 
At the initial time $t=0$, $\shw$ starts from unity,
which reflects that the initial system is 
given by a coherent state.
One sees that $\shw$ first increases rapidly, 
then shows only a gradual increase, and finally 
reaches some value and hardly changes. 
The saturation of $\shw$ indicates that the system reaches the thermal equilibrium.
The final value of $\shw$ may be the HW entropy in the thermal equilibrium 
in the classical limit, which is estimated as 
\begin{align}
\shw^{\rm cl.eq.}\sim 3.1\ .\label{Eq:cleq}
\end{align}

To compare with the thermal HW entropy shown in Eq.~\refeq{Eq:cleq}, 
we consider the classical Gibbs ensemble of the free field, 
\begin{align}
\fw^{\rm cl.eq.}(\{\Phik,\Pik\})|_{\lambda=0}\propto e^{-H_{\rm free}/T}
=\prod_{\bk} \exp{\left[-\frac{\omega_{\bk}^2\Phi^2_{\bk}+\Pi^2_{\bk}}{2T}\right]}\ ,
\end{align}
which gives thermal expectation values of observables in the classical limit at $\lambda=0$. 
Then, we get the Husimi function,
\begin{align}
\fh^{\rm cl.eq.}(\{\Phik,\Pik,\omegak\})|_{\lambda=0}\propto\prod_{\bk} \exp{\left[-\frac{\omega_{\bk}^2\Phi^2_{\bk}+\Pi^2_{\bk}}{2(T+\omega_{\bk}/2)}\right]}\ ,
\end{align}
and the HW entropy,
\begin{align}
\shw^{\rm cl.eq.}|_{\lambda=0}=\sum_{\bk} \left[ 1 + \ln{\left(\frac{T+\omega_{\bk}/2}{\omega_{\bk}}\right)} \right]\ .\label{Eq:HWfree}
\end{align}
%
Assuming that the system is in the thermal equilibrium
after the HW entropy stops to increase, 
we extract the temperature of the system through the equipartition relation $\average{\pi(x)^2}=T$ 
that holds in thermal equilibrium in the classical limit. 
By substituting the numerically extracted temperature $T=20$ into Eq.~\refeq{Eq:HWfree},
we obtain $\shw^{\rm cl.eq.}|_{\lambda=0}$ at the same temperature as our simulation, 
$\shw^{\rm cl.eq.}\Big|_{\lambda=0,T=20} \sim 3.2$.
This theoretical estimate 
and our numerical estimation $\shw^{\rm cl.eq.}\Big|_{\lambda=1,T=20}\sim 3.1$ in Eq.~\refeq{Eq:cleq} 
agree well with each other,
and it indicates that our simulation parameter at $\lambda = 1$
describes the relaxation to thermal equilibrium in the weak coupling region of the theory.
Note that since the HW entropy shown in Eq.~\refeq{Eq:HWfree} is based on the classical thermal equilibrium distribution, 
its entropy density has the UV divergence in the continuum limit.
To circumvent this problem, it is useful to use a framework that can perturbatively take account of higher-order quantum effects, 
including quantum statistical properties, such as the kinetic theory~\cite{AMYkinetic} 
and the two-particle irreducible (2PI) effective action approach~\cite{Berges:2004yj,Aarts:2001yn,Hatta-Nishiyama}.
Aside from these descriptions, several improved classical field methods have been proposed to manage the UV divergence:
introducing counter terms~\cite{Aarts}, integrating high-momentum modes assumed to be a heat bath~\cite{Bodeker:1995pp,Greiner:1996dx}, 
taking account of the explicit coupling of fields and particles~\cite{kineticfield}, 
and considering a classical field theory with quantum statistical nature~\cite{replica}.

\section{Husimi-Wehrl entropy of the Yang-Mills field in expanding geometry}\label{Sec:YM}
Here we investigate the dynamical production of the HW entropy in the
semiclassical time evolution of the SU($2$) Yang-Mills field in the expanding geometry. 
We also compute the 
pressure and particle number, and discuss the relation between the HW entropy and them. 
We make all the quantities dimensionless normalizing with the transverse spatial lattice spacing $a$, 
and the $\eta$ direction gauge field $A_\eta$ is normalized by the spatial lattice spacing $a_\eta$ in $\eta$ direction.
\subsection{formulation}\label{Sec:YM_theory}
The evolution of the longitudinally expanding system is discussed using the $\tau$-$\eta$ coordinate, 
$\left( x^\tau,\ x^1,\ x^2,\ x^\eta \right) = \left( \tau=\sqrt{t^2-z^2},\ x,\ y,\ \eta=\frac{1}{2}\ln{\frac{t+z}{t-z} }\right)$, 
with the metric, $g_{\mu\nu}={\rm diag}(1,-1,-1,-\tau^2)$. 
We consider the non-compact Hamiltonian in the Fock-Schwinger gauge $A_\tau=0$ 
on a $L^2_\perp \times L_\eta$ lattice, 
\begin{align}
H = \frac{1}{2a_\eta\tau} \sum_{a,\bm{x}} \left[ \left(\hat{E}^{a1}(x)^2 + \hat{E}^{a2}(x)^2 + \hat{B}^{a1}(x)^2 + \hat{B}^{a2}(x)^2 \right) 
+ (a_\eta\tau)^2\left( \hat{E}^{a\eta}(x)^2 + \hat{B}^{a\eta}(x)^2 \right)\right]\ .\label{Eq:YM_ham}
\end{align}
The color electric and color magnetic fields are defined as
\begin{align}
&\hat{E}^{ai}(x)=\al\tau\partial_\tau \hat{A}^a_i(x)\ \ (i=1,2)\ ,\ \ \hat{E}^{a\eta}(x)=\frac{1}{\al\tau}\partial_\tau \hat{A}^a_\eta(x)\ ,\\
&\hat{B}^{ai}(x)=\frac{1}{2}\epsilon^{ijk} \left[\pf_j \hat{A}^a_k(x)-\pf_k \hat{A}^a_j(x)  
- \frac{g^2}{4}f^{abc}\left(\hat{A}^b_j(x)+\hat{A}^b_j(x+\hat{k})\right)\left(\hat{A}^c_k(x)+\hat{A}^c_k(x+\hat{j})\right)\right]\nonumber\\
&\ \ \ \ \ \ \ \ \ \ (i=1,2,\eta)\ ,
\end{align}
where $\hat{i}$ is the unit vector in the $i$-direction on the lattice. 
Here we show the expression of the gauge field in terms of the annihilation and creation operators 
for the free Yang-Mills field~\cite{Berges2014_2,flu_EG}, which is shown in App.~\ref{App:sc_exp}, 
\begin{align}
&\hat{A}^a_\mu(x)=\frac{1}{\sqrt{L^2_\perp L_\eta}}\sum_{\lambda,\bk}  
\left(  \hat{a}^{a\lambda}_{\bk}\tilde{\mathcal{A}}^\lambda_{\mu,\bk}(\tau) e^{i(\bk_\perp\cdot\bx_\perp+\nu\eta)} + {\rm h.c.}\right)\ ,\\
&\tilde{\mathcal{A}}^1_{\mu,\bk}(\tau)=\frac{i}{\omega_{\bm{k}_\perp}}\sqrt{\frac{\pi}{4\al}}e^{\frac{\pi |\tilde{\nu}|}{2\al}}
(0, \tilde{k}_2, \tilde{k}_1, 0)H^{(2)}_{i|\tilde{\nu}|/\al}(\omega_{\bm{k}_\perp}\tau)\ ,\\
&\tilde{\mathcal{A}}^2_{\mu,\bk}(\tau)=-\frac{\tilde{\nu}^*}{\al \omega_{\bm{k}_\perp}}\sqrt{\frac{\pi}{4\al}}e^{\frac{\pi |\tilde{\nu}|}{2\al}}
\left(0, \tilde{k}_1\alpha^{(2)}_{i|\tilde{\nu}|/\al}(\omega_{\bm{k}_\perp}\tau), \tilde{k}_2\alpha^{(2)}_{i|\tilde{\nu}|/\al}(\omega_{\bm{k}_\perp}\tau), 
-\frac{\al^2}{\tilde{\nu}^*}\beta^{(2)}_{i|\tilde{\nu}|/\al}(\omega_{\bm{k}_\perp}\tau) \right)\ ,\\
&\alpha^{(2)}_{i|\tilde{\nu}|/\al}(\omega_{\bm{k}_\perp}\tau)=\int^{\omega_{\bm{k}_\perp}\tau}_{\omega_{\bm{k}_\perp}\tau_0} dz 
 \frac{1}{z}H^{(2)}_{i|\tilde{\nu}|/\al}(z)
-\frac{ \omega_{\bm{k}_\perp}\tau_0   }{(|\tilde{\nu}|/\al)^2+(\omega_{\bm{k}_\perp}\tau_0)^2}\dot{H}^{(2)}_{i|\tilde{\nu}|/\al}(\omega_{\bm{k}_\perp}\tau_0)\ ,\\
&\beta ^{(2)}_{i|\tilde{\nu}|/\al}(\omega_{\bm{k}_\perp}\tau)=\int^{\omega_{\bm{k}_\perp}\tau}_{\omega_{\bm{k}_\perp}\tau_0} dz
          z H^{(2)}_{i|\tilde{\nu}|/\al}(z)
-\frac{(\omega_{\bm{k}_\perp}\tau_0)^3}{(|\tilde{\nu}|/\al)^2+(\omega_{\bm{k}_\perp}\tau_0)^2}\dot{H}^{(2)}_{i|\tilde{\nu}|/\al}(\omega_{\bm{k}_\perp}\tau_0)\ ,
\end{align}
where $\omega_{\bm{k}_\perp}=2\sqrt{\sin^2{(\frac{k_1}{2})}+\sin^2{(\frac{k_2}{2})}}$ is a transverse frequency on the lattice,
$\tilde{k}_i=2e^{i\frac{k_i}{2}}\sin{\frac{k_i}{2}}\ \ (i=1,2)$ and $\tilde{\nu}=2e^{i\frac{\nu}{2}}\sin{\frac{\nu}{2}}$ are 
the discrete Fourier transforms of the forward difference operators 
in the $i$-direction and $\eta$-direction, $\partial^{\rm F}_i$ and $\partial^{\rm F}_\eta$, 
on the lattice, $H^{(2)}$ is the Hankel function of the second kind and $\tau_0$ is the initial proper time in our simulations.
Here the residual gauge degrees of freedom are fixed by the Coulomb type gauge condition, 
$\left( \partial_1 A_1 + \partial_2 A_2 + \tau^{-2}\partial_\eta A_\eta \right)\Bigl|_{\tau=\tau_0}=0$~\cite{Berges2014_2}.
Then the electric field, $\hat{E}$, and the free part of the magnetic field, $\hat{B}_0$, can be also expressed 
in terms of ${\hat{a}^{a\lambda}_{\bk}}$, 
\begin{align}
&\hat{E}^{ai}(x)  =\frac{1}{\sqrt{L^2_\perp L_\eta}}\sum_{\lambda,\bk} 
                   \left( \hca^{a\lambda}_{\bk}\mathcal{E}^{\lambda, i}_{\bk}(\tau) e^{i(\bk_\perp\cdot\bx_\perp+\nu \eta)} + {\rm h.c.}\right)\ ,\label{Eq:sq_e2}\\
&\hat{B}^{ai}_0(x)=\epsilon^{ijk} \partial^{\rm F}_j A_k(x)
                  =\frac{1}{\sqrt{L^2_\perp L_\eta}}\sum_{\lambda,\bk} 
                   \left( \hca^{a\lambda}_{\bk}\mathcal{B}^{\lambda, i}_{\bk}(\tau) e^{i(\bk_\perp\cdot\bx_\perp+\nu \eta)} + {\rm h.c.}\right)\ ,\label{Eq:sq_b2}\\
&\mathcal{E}^{1,i}_{\bk}(\tau)=\dot{H}^{(2)}_{i|\tilde{\nu}|/\al}(\omega_{\bm{k}_\perp}\tau)\varepsilon^{1,i }_{\bk}\ ,\ \ 
 \mathcal{E}^{2,i}_{\bk}(\tau)=     H ^{(2)}_{i|\tilde{\nu}|/\al}(\omega_{\bm{k}_\perp}\tau)\varepsilon^{2,i }_{\bk}\ ,\\
&\mathcal{B}^{1,i}_{\bk}(\tau)=     H ^{(2)}_{i|\tilde{\nu}|/\al}(\omega_{\bm{k}_\perp}\tau)\varepsilon^{2,i*}_{\bk}\ ,\ \ 
 \mathcal{B}^{2,i}_{\bk}(\tau)=\dot{H}^{(2)}_{i|\tilde{\nu}|/\al}(\omega_{\bm{k}_\perp}\tau)\varepsilon^{1,i*}_{\bk}\ ,\\
&\varepsilon^{1,i}_{\bk} =i\al\tau\sqrt{\frac{\pi}{4\al}}e^{\frac{\pi |\tilde{\nu}|}{2\al}}(\tilde{k}^*_2, -\tilde{k}^*_1, 0)\ ,\ \ 
 \varepsilon^{2,i}_{\bk} =- \frac{\tilde{\nu}^*}{\omega_{\bm{k}_\perp}}\sqrt{\frac{\pi}{4\al}}e^{\frac{\pi |\tilde{\nu}|}{2\al}}
 \left(\tilde{k}_1, \tilde{k}_2, -\frac{\omega^2_{\bm{k}_\perp}}{\tilde{\nu}^*} \right)\ .
\end{align}

By substituting Eq.~\refeq{Eq:sq_e2} and Eq.~\refeq{Eq:sq_b2} for the electric field and magnetic field in Eq.~\refeq{Eq:YM_ham} respectively, 
we can obtain the free part of the Hamiltonian as 
\begin{align}
&H_{\rm free}
=\sum_{a,\lambda,\bk} |\Delta_{\bk}| \left(\hca^{a\lambda\dagger}_{\bk}\hca^{a\lambda}_{\bk} + \frac{1}{2}\right) 
 +\frac{1}{2} \sum_{a,\lambda,\bk} \left(\Delta_{\bk}\hca^{a\lambda}_{\bk}\hca^{a\lambda}_{-\bk}
 +\Delta^*_{\bk} \hca^{a\lambda\dagger}_{\bk}\hca^{a\lambda\dagger}_{-\bk} \right)\ ,\label{Eq:Hfree}\\
&\Delta_{\bk}
=\frac{\pi\tau }{4}e^{\frac{\pi |\tilde{\nu}|}{\al}}
 \left( (\omega^2_{\bm{k}_\perp}+|\tilde{\nu}/(\tau\al)|^2) \left\{     H ^{(2)}_{i|\tilde{\nu}|/\al}(\omega_{\bm{k}_\perp}\tau)\right\}^2
      + (\omega_{\bm{k}_\perp})^2 \left\{\dot{H}^{(2)}_{i|\tilde{\nu}|/\al}(\omega_{\bm{k}_\perp}\tau)\right\}^2 \right)\ .
\end{align}
The first term of Eq.~\refeq{Eq:Hfree} is asymptotically regarded as a set of harmonic oscillators,
\begin{align}
\sum_{a,\lambda,\bk} |\Delta_{\bk}| \hca^{a\lambda\dagger}_{\bk} \hca^{a\lambda}_{\bk}
\longrightarrow \sum_{a,\lambda,\bk} \omega_{\bm{k}_\perp} \hca^{a\lambda\dagger}_{\bk} \hca^{a\lambda}_{\bk}\ .
\end{align}
and thus we can define canonical variables that are asymptotically regarded as harmonic oscillators, 
\begin{align}
\hPhi^{a\lambda}_{\bk} &\equiv  \frac{1}{\sqrt{2\omega_{\bm{k}_\perp}}}  \left[ \hca^{a\lambda}_{\bk}+\hca^{a\lambda\dagger}_{\bk}\right]\ ,\label{Eq:phidot2}\\
\hPi ^{a\lambda}_{\bk} &\equiv  -i\sqrt{\frac{\omega_{\bm{k}_\perp}}{2}} \left[ \hca^{a\lambda}_{\bk}-\hca^{a\lambda\dagger}_{\bk}\right]\ .\label{Eq:pidot2}
\end{align}

In the same way as the scalar field, we consider the Wigner function for newly defined canonical 
variables $\{\hPhi,\hPi\}$, whose initial condition is given by 
the product of coherent states of each mode, 
$|\{\alpha^{a,\lambda}_{\bk}\}\rangle=\prod_{a,\lambda,\bk}|\alpha^{a,\lambda}_{\bk}\rangle$,
\begin{align}
\fw(\{\Phi,\Pi\})
&=\prod_{a,\lambda,\bk} \int {\rm d}\Phi'^{a\lambda}_{\bk}
  \average{\Phi^{a\lambda}_{\bk}+\frac{\Phi'^{a\lambda}_{\bk}}{2}\,\Big|\,\alpha^{a\lambda}_{\bk}}
  \average{\alpha^{a\lambda}_{\bk}\,\Big|\,\Phi^{a\lambda}_{\bk}-\frac{\Phi'^{a\lambda}_{\bk}}{2}}e^{-i\Pi^{a\lambda}_{\bk}\Phi'^{a\lambda}_{\bk}}\ ,\label{Eq:WiginiCYM}
\end{align}
We also consider the Husimi function and HW entropy for $\{\hat{\Phi},\hat{\Pi}\}$ with smearing parameters that 
are taken as the same values as asymptotic eigenfrequencies $\{\omega_{\bm{k}_\perp}\}$, 
$\fh(\{\Phi,\Pi,\omega_\perp\})$ and $\shw(\{\omega_\perp\})$. 
In the later discussion of this section, we omit $\{\omega_\perp\}$ in the expression of the HW 
entropy of the Yang-Mills field: $\shw$. 

\subsection{Numerical results}
Here we investigate the semiclassical evolution of the pressure, the HW entropy and the distribution 
function of particles in the Yang-Mills theory on the $32^2\times 420$ lattice with periodic boundary 
condition.
The coupling constant is taken as $g=0.1$ and $0.2$ and the longitudinal size of the system is taken 
as $V_\eta(=L_\eta \times a_\eta)=2$.
The test particle positions in the phase space at $\tau=\tau_0$ are generated randomly according to 
the initial Wigner function shown in Eq.~\refeq{Eq:WiginiCYM}.
It is shown in the next paragraph how to give the initial macroscopic fields 
$({\bar\Phi^{a\lambda}_{\bk}},{\bar\Pi^{a\lambda}_{\bk}})$.
The classical equation of motion for each test particle is solved by leapfrog integration. 
The number of test particles is taken as $\NTP=96$. 

We prepare the initial macroscopic fields so as to mimic the the glasma initial 
condition~\cite{MV1994_1,MV1994_2,MV1994_3,LMW1995_1,LMW1995_2} as follows; 
the color electric and color magnetic fields are boost invariant and parallel to the collision axis 
as, 
\begin{align}
&\left|\average{\hat{a}^{a\lambda}_{\bk}}\right|^2\Big|_{\tau=\tau_0}
=\delta_{a,1} \left(\frac{V_\eta}{2\pi}\delta_{\nu,0}\right) \frac{\Delta}{\alpha_s}\cdot 
\left|f(\omega_{\bm{k}_\perp})\right|^2\ ,\label{Eq:ini11}\\
&\average{\hat{a}^{a\lambda}_{\bk}}\Big|_{\tau=\tau_0}
=\average{\hat{a}^{a\lambda\dagger}_{-\bk}}\Big|_{\tau=\tau_0}\ ,\label{Eq:ini22} 
\end{align}
where $\Delta$ is an arbitrary free parameter and $f(\omega_{\bm{k}_\perp})$ is the transverse 
momentum distribution. 
The boost invariance is guaranteed by $\frac{V_\eta}{2\pi}\delta_{\nu,0}$, which is the lattice
representation of the delta function $\delta(\nu)$.
The phase of $\average{\hat{a}^{a\lambda}_{\bk}}$ varies randomly. 
Since the direction of $\average{\hat{a}^{a\lambda}_{\bk}}$ 
in the color space is toward the $1$ direction, the gauge field is also directed toward
the $1$ direction in the color space, 
$\average{\hat{A}^{\lambda}_{\bk}} \propto \delta_{a,1}$. 
Thus, the initial macroscopic magnetic field turns out to be equal to its free part shown 
in Eq.~\refeq{Eq:sq_b2}, 
$\average{\hat{B}}\Big|_{\tau=\tau_0}=\average{\hat{B}_0}\Big|_{\tau=\tau_0}$.
Here, we assume that all the contribution of fluctuations in 
$\average{\hat{B}}\Big|_{\tau=\tau_0}$ is subtracted 
as the vacuum contribution or other divergences.
Then, by substituting Eq.~\refeq{Eq:ini11} into Eq.~\refeq{Eq:sq_b2} and Eq.~\refeq{Eq:sq_e2} 
and utilizing Eq.~\refeq{Eq:ini22}, 
we obtain the analytic expression of nonzero parts of $\average{\hat{\bm{B}}_{\bk}}$ and 
$\average{\hat{\bm{E}}_{\bk}}$ as 
\begin{align}
&\average{\hat{E}^{a1}   _{\bk}}\Big|_{a=1,\nu=0}=\average{\hat{a}^{1\lambda}_{\bk}}\Big|_{\tau=\tau_0}\cdot
(i\al\tau_0\tilde{k}^*_2)   \sqrt{\frac{\pi}{4 \al}} {\rm Re}\left(\dot{H}^{(2)}_0(\omega_{\bm{k}_\perp}\tau_0)\right)\ ,\label{mac:e1}\\
&\average{\hat{E}^{a2}   _{\bk}}\Big|_{a=1,\nu=0}=\average{\hat{a}^{1\lambda}_{\bk}}\Big|_{\tau=\tau_0}\cdot
(i\al\tau_0\tilde{k}^*_1)   \sqrt{\frac{\pi}{4 \al}} {\rm Re}\left(\dot{H}^{(2)}_0(\omega_{\bm{k}_\perp}\tau_0)\right)\ ,\label{mac:e2}\\
&\average{\hat{E}^{a\eta}_{\bk}}\Big|_{a=1,\nu=0}=\average{\hat{a}^{2\lambda}_{\bk}}\Big|_{\tau=\tau_0}\cdot
\omega_{\bm{k}_\perp}                \sqrt{\frac{\pi}{4 \al}} {\rm Re}\left(     H ^{(2)}_0(\omega_{\bm{k}_\perp}\tau_0)\right)\ ,\label{mac:e3}\\
&\average{\hat{B}^{a1}   _{\bk}}\Big|_{a=1,\nu=0}=\average{\hat{a}^{2\lambda}_{\bk}}\Big|_{\tau=\tau_0}\cdot
(i\al\tau_0\tilde{k}^*_2)^* \sqrt{\frac{\pi}{4 \al}} {\rm Re}\left(\dot{H}^{(2)}_0(\omega_{\bm{k}_\perp}\tau_0)\right)\ ,\label{mac:b1}\\
&\average{\hat{B}^{a2}   _{\bk}}\Big|_{a=1,\nu=0}=\average{\hat{a}^{2\lambda}_{\bk}}\Big|_{\tau=\tau_0}\cdot
(i\al\tau_0\tilde{k}^*_1)^* \sqrt{\frac{\pi}{4 \al}} {\rm Re}\left(\dot{H}^{(2)}_0(\omega_{\bm{k}_\perp}\tau_0)\right)\ ,\label{mac:b2}\\
&\average{\hat{B}^{a\eta}_{\bk}}\Big|_{a=1,\nu=0}=\average{\hat{a}^{1\lambda}_{\bk}}\Big|_{\tau=\tau_0}\cdot
\omega_{\bm{k}_\perp}                \sqrt{\frac{\pi}{4 \al}} {\rm Re}\left(     H ^{(2)}_0(\omega_{\bm{k}_\perp}\tau_0)\right)\ .\label{mac:b3}
\end{align}

On account of the formula,
\[
{\rm Re}H^{(2)}_0(z)=1-\frac{z^2}{4}+\mathcal{O}(z^3),
\] 
the transverse components of $\average{\hat{\bm{B}}_{\bk}}$ and 
$\average{\hat{\bm{E}}_{\bk}}$ are proportional to $(\omega_{\bm{k}_\perp} \tau_0)^2$
when $\omega_{\bm{k}_\perp} \tau_0 \ll 1$, while the longitudinal components 
of $\average{\hat{\bm{B}}_{\bk}}$ and $\average{\hat{\bm{E}}_{\bk}}$ 
are independent of $\omega_{\bm{k}_\perp} \tau_0$ in this limit.
Therefore, when $\omega_{\bm{k}_\perp} \tau_0 \ll 1$, the transverse components are much smaller than the longitudinal components.

We consider two different profiles of the initial macroscopic fields, 
which are distinguished from each other by 
different momentum distributions $f(\omega_{\bm{k}_\perp})$: 
One $f(\omega_{\bm{k}_\perp})$ is given by a step function, while the other is given by 
a difference of Gaussians as adopted in \cite{Kovini}, 
\begin{align}
&{\rm Type\ 1}\ \ |f_1(\omega_{\bm{k}_\perp})|^2=\Theta(Q_s-\omega_{\bm{k}_\perp})\ ,\label{Eq:ini1}\\
&{\rm Type\ 2}\ \ |f_2(\omega_{\bm{k}_\perp})|^2=\frac{Q^2_s}{2\omega^2_{\bm{k}_\perp}}
\left[ e^{-k^2_\perp/2Q^2_s} - e^{-k^2_\perp/Q^2_s} \right]\ ,\label{Eq:ini2}
\end{align}
where $Q_s$ characterizes the typical transverse momentum. 
We take $Q_s$ as $Q_s\tau_0=0.1 \ll 1$, 
which implies that the macroscopic color electric and color magnetic 
fields are parallel to the collision axis.

\subsubsection{Pressure isotropization}
\ \\

The energy-momentum (EM) tensor is defined as
\begin{align}
\hat{T}_{\mu\nu}(x)\equiv 
-g^{\kappa\sigma}\hat{F}^a_{\mu\kappa}(x)\hat{F}^a_{\nu\sigma}(x)
+\frac{1}{4}g_{\mu\nu}g^{\alpha\beta}g^{\gamma\omega}\hat{F}^a_{\alpha\gamma}(x)\hat{F}^a_{\beta\omega}(x)\ .
\end{align}
To define the pressure and the energy density in the expanding geometry, 
two types of subtraction are necessary~\cite{PressureEG2013}, 
\begin{align}
&\varepsilon \equiv \frac{1}{V}\sum_x g_{\tau\tau} \average{\hat{T}^{\tau\tau}(x)}_{\rm mac+fluc} - \frac{1}{V}\sum_x g_{\tau\tau} \average{\hat{T}^{\tau\tau}(x)}_{\rm vac} -                \frac{\alpha}{\tau^2}\ ,\\
&P_i         \equiv \frac{1}{V}\sum_x g_{ii      } \average{\hat{T}^{ii      }(x)}_{\rm mac+fluc} - \frac{1}{V}\sum_x g_{ii      } \average{\hat{T}^{ii      }(x)}_{\rm vac} - \delta_{i,\eta}\frac{\alpha}{\tau^2}\ ,
\end{align}
where $\average{}_{\rm mac+fluc}$ denotes the average in the test particles with the initial conditions 
given in Eq.~\refeq{Eq:ini1} or Eq.~\refeq{Eq:ini2}, $\average{}_{\rm vac}$ denotes the vacuum 
contribution, and $\alpha/\tau^2$ is the remaining divergence after subtracting the vacuum 
contribution;
see App.~\ref{App:ini_div} for the details.

In actual calculations,  the vacuum contribution $\average{\hat{T}^{\mu\mu}}_{\rm vac}$ is evaluated 
by the test particle method with the vanishing initial macroscopic fields, 
$\average{\hat{\Phi}^{a\lambda}_{\bk}}\Big|_{\tau=\tau_0}=\average{\hat{\Pi}^{a\lambda}_{\bk}}\Big|_{\tau=\tau_0}=0$. 
Following Ref.~\cite{PressureEG2013},
the remaining divergent part $\alpha/\tau^2$ is extracted phenomenologically
by a fitting procedure as shown in App.~\ref{App:ini_div}.

\begin{figure}[hbtp]
\centering{
\includegraphics[width=80mm,bb=0 0 360 252]{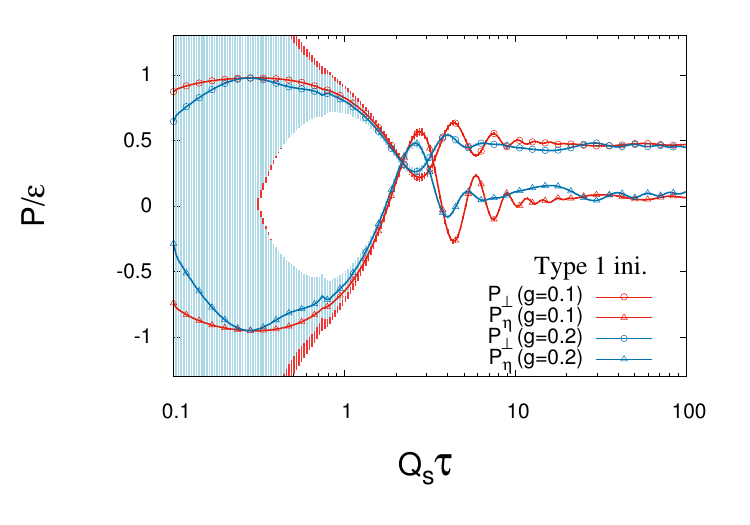}%
\includegraphics[width=80mm,bb=0 0 360 252]{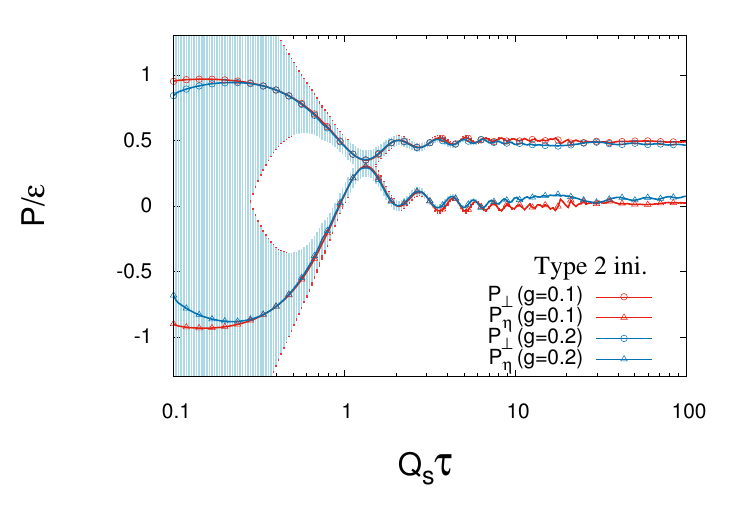}%
}\caption{
The time evolution of the transverse and longitudinal pressures, $P_\perp$ and $P_\eta$, 
per energy density $\varepsilon$ of Yang-Mills theory in the expanding geometry 
on the $32^2 \times 420$ lattice, with $g=0.1$ and $0.2$; the smearing parameter $\sigma$ 
is set by a constraint $\sigma/g^2=1000$.
The left and right panels show the results 
with the  initial condition of type 1 and 2, respectively.
The number of test particles is 
$\NTP=96$. 
The 
lines with circles and triangles show $P_\perp/\varepsilon$ and 
$P_\eta/\varepsilon$, respectively.
}
\label{Fig:pressure}
\end{figure}

Figure~\ref{Fig:pressure} shows the time evolution of the transverse and longitudinal pressures, 
$P_\perp$ and $P_\eta$, normalized by the energy density $\varepsilon$ at $g=0.1$ and $0.2$ 
for the two types of initial conditions.  
It is noted that error bars are large at small $Q_s\tau$ 
because large subtractions are needed. 
In the initial stage with $Q_s\tau<(1-2)$, $P_\eta$ is negative, which
reflects that the macroscopic color electric and magnetic fields are parallel to the collision 
axis in the earliest stage.
Then till $Q_s\tau<(2-3)$, $P_\perp$ and $P_\eta$ tend to come closer, and there is no 
$g$ dependence of their values at this period.
In the later stage with $Q_s\tau>(2-3)$, $P_\perp/\varepsilon$ and $P_\eta/\varepsilon$
gradually approach some different constant values, respectively, with oscillatory behaviors, 
the amplitudes of which become tiny in the large $Q_s \tau$ region. The ratio $P_\eta/P_\perp$ 
in the final stage clearly deviates from unity;
although it slightly gets 
closer for larger $g$ $(=0.2)$ the difference between them is still large, 
which means that the isotropization of the pressure is not achieved with $g\leq 0.2$.
In a previous study based on the McLerran-Venugopalan model~\cite{PressureEG2013}, 
such pressure isotropization was found at $g=0.5$ and not at $g=0.1$. 
Therefore, our results do not contradict their results.
It is noted that the nearly constant behavior of 
$P_\perp/P_\eta$ at later times shown in Fig.~\ref{Fig:pressure}
seems consistent with the kinetics results~\cite{Kurkela}. 
Such consistency may imply that the semiclassical description 
and the kinetic description commonly take into account one of the essential processes 
in the Yang-Mills theory~\cite{MSkinetic,Jkinetic}.

\subsubsection{Creation and growth of (Husimi-Wehrl) entropy}
\ \\
\begin{figure}[htbp]
\begin{minipage}{0.5\hsize}
\centering\includegraphics[width=80mm,bb=0 0 360 252]{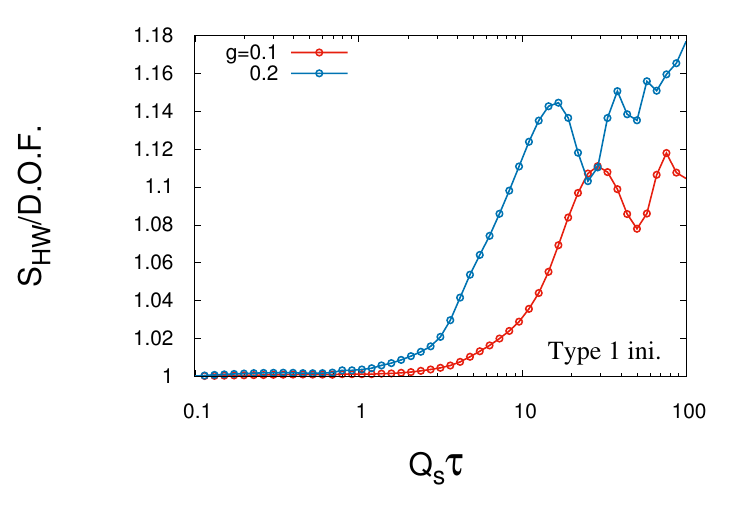}\\
\end{minipage}
\begin{minipage}{0.5\hsize}
\centering\includegraphics[width=80mm,bb=0 0 360 252]{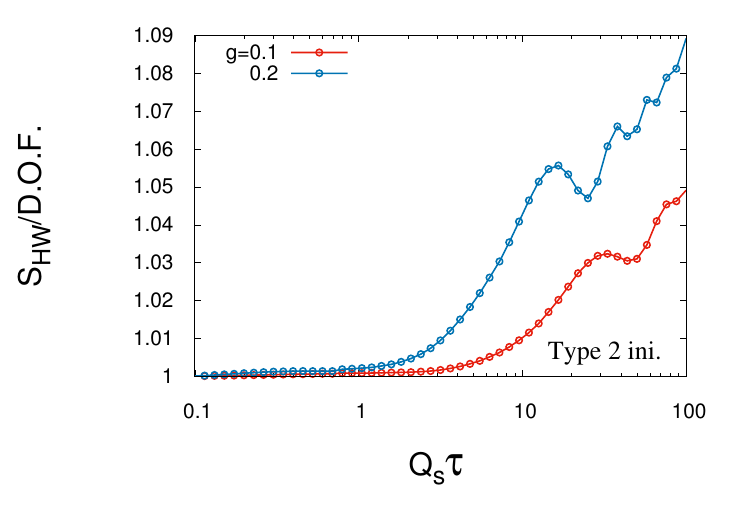}\\
\end{minipage}
\caption{The time evolution of the HW entropy $\shw$ per degrees of freedom 
with the initial condition of type 1(left panel) and type 2(right panel).
The other calculational conditions are the same as in
Fig.~\ref{Fig:pressure}. 
}
\label{Fig:entropyYM}
\end{figure}

In Fig.~\ref{Fig:entropyYM}, we show the time evolution of $\shw=\shwa$ per degrees of freedom 
with $g=0.1$ and $0.2$ for the two types of initial condition. 
Firstly, we remark
that $\shw$ initially agrees with unity with an error of less than 0.01\%, which is in accordance 
with the fact that the initial conditions are prepared as a coherent state. 
In the earliest stage with $Q_s\tau<(2-3)$, the HW entropy hardly increases, and then in the 
intermediate stage with $Q_s\tau<(20-30)$, it shows a rapid growth.
Then it still does show an increases on average but with smaller growth rate and 
an oscillatory behavior imposed.
For both of the initial conditions, the larger the coupling constant, the larger the growth rate of 
the HW entropy. 
In Ref.~\cite{Tsukiji2016,Tsukiji2018}, it was shown in the semiclassical simulation with the 
{\it non-expanding geometry} that  in the last stage where the HW entropy production has 
been saturated with a small production speed, the Yang-Mills field configuration is already close to 
that in equilibrium.
It should be noticed, however, that the similar slow production rate of the HW entropy seen in
Fig.~\ref{Fig:entropyYM} does not readily mean that the system is near equilibrium 
since the large anisotropy of the pressure still remains in the present case with an expanding 
geometry, which may account for, at least partly, the slow production rate of the entropy which is 
actually caused by a fact that the system is still in a nonequilibrium state.

\subsubsection{Relationship of particle distribution and Husimi-Wehrl entropy}
\ \\

\begin{figure}[htbp]
\centering{
\includegraphics[width=80mm,bb=0 0 360 252]{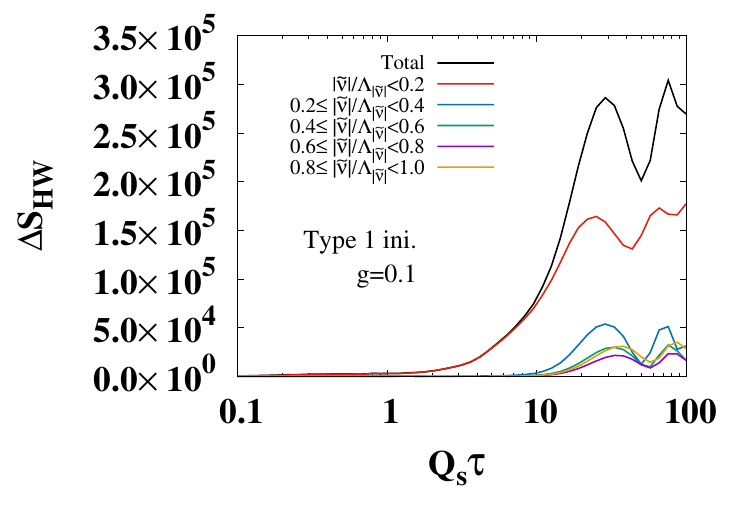}%
\includegraphics[width=80mm,bb=0 0 360 252]{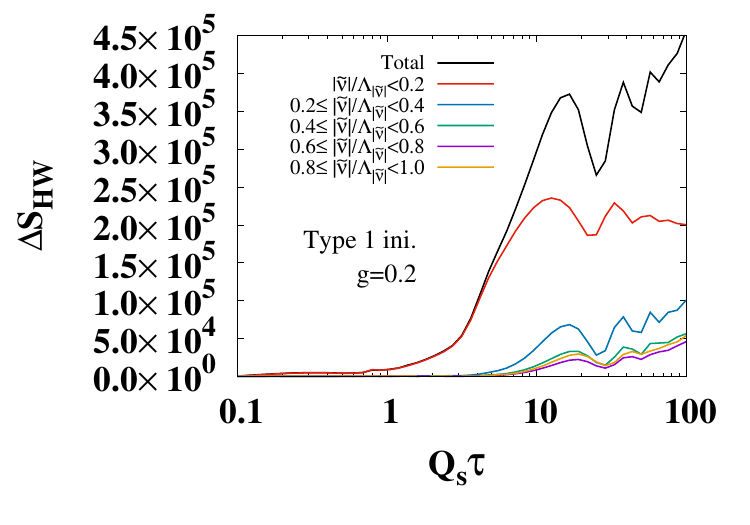}%
\\
\includegraphics[width=80mm,bb=0 0 360 252]{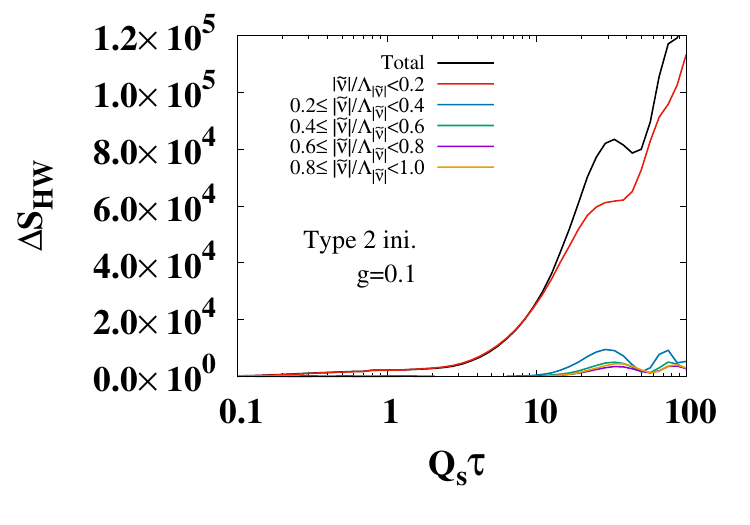}%
\includegraphics[width=80mm,bb=0 0 360 252]{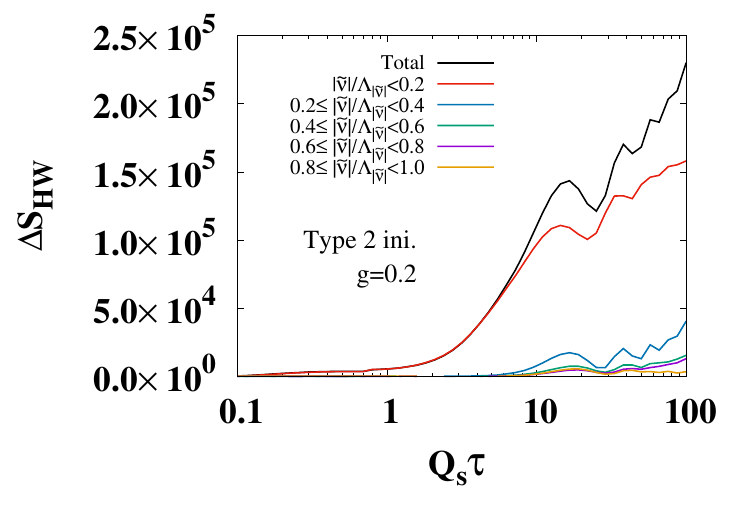}%
}
\caption{The HW entropy increase after time evolution,
$\Delta \shw = \shw-N_{\rm dof}$,
in several longitudinal momentum intervals with $g=0.1$ and $0.2$ 
for the two types of initial condition.
The bin size in the longitudinal momentum is 
$|\tilde{\nu}|/\Lambda_{|\tilde{\nu}|}=0.2$.
The left (right) panels show the numerical results with $g=0.1 (0.2)$,
and upper (lower) panels are for the type 1 (type 2) initial condition.
}
\label{Fig:entropy_long}
\end{figure}

To understand the underlying mechanism of the HW entropy production,
we investigate the time evolution of the particle number 
as well as the HW entropy in {\it piecewise} with respect
to different longitudinal momentum modes with an interval 
$|\tilde{\nu}|/\Lambda_{|\tilde{\nu}|}=0.2$,
where $\Lambda_{|\tilde{\nu}|}$ denotes the ultraviolet cutoff
of the momentum.
Figure~\ref{Fig:entropy_long} shows the entropy increase
after time evolution $\Delta \shw = \shw-N_{\rm dof}$,
in several longitudinal momentum intervals with $g=0.1$ and $0.2$
for the two types of initial conditions.
We find that the large HW entropy is first produced
in the lowest longitudinal momentum interval, 
$|\tilde{\nu}|/\Lambda_{|\tilde{\nu}|}<0.2$,
which is followed by an slow increase of the HW entropy 
in the higher longitudinal momentum intervals.
Thus, the rapid production of the HW entropy shown 
in Fig.~\ref{Fig:entropyYM} occurs in the low longitudinal momentum modes.

\begin{figure}[htbp]
\centering{
\includegraphics[width=80mm,bb=0 0 360 252]{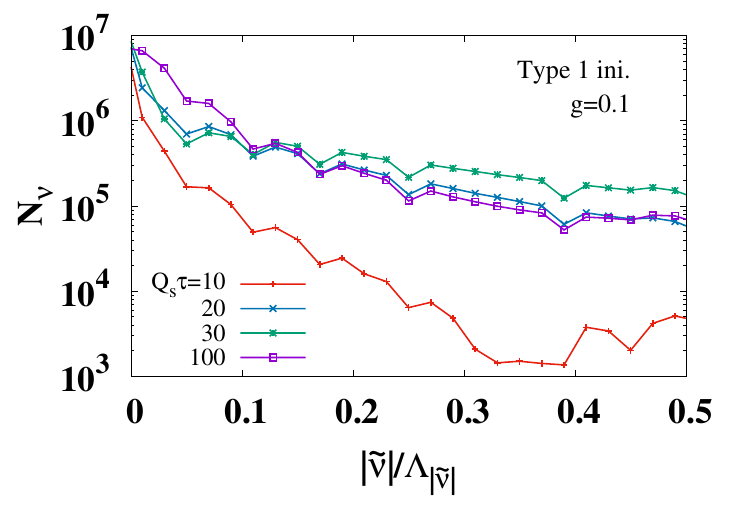}%
\includegraphics[width=80mm,bb=0 0 360 252]{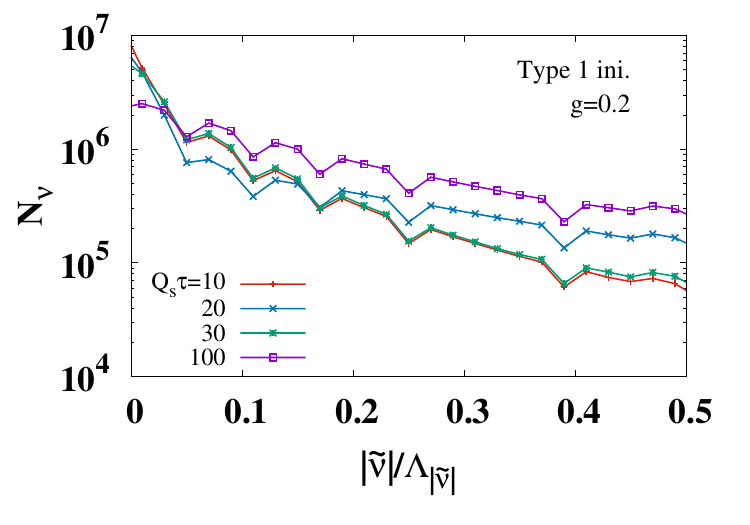}%
\\
\includegraphics[width=80mm,bb=0 0 360 252]{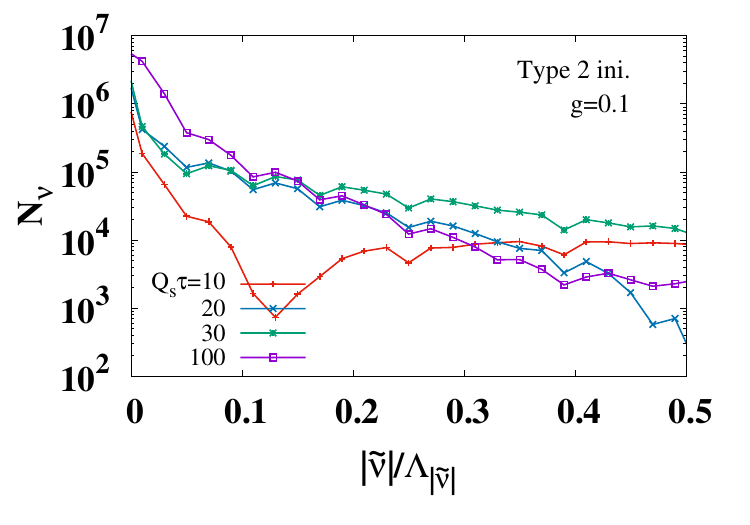}%
\includegraphics[width=80mm,bb=0 0 360 252]{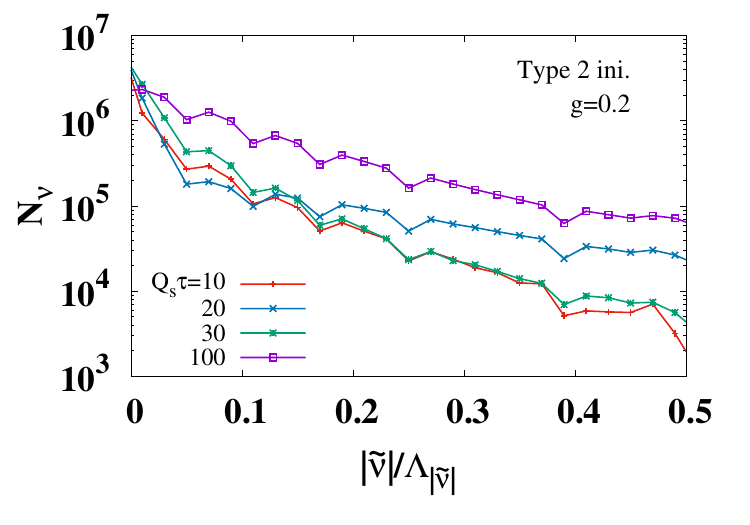}%
}
\caption{The time evolution of the effective particle-number distribution
$N_{\nu}$ as a function of the longitudinal momentum $\tilde{\nu}$
as defined in \refeq{eq:eff-dis-number-long}
with the two initial conditions and coupling constants $g=0.1$ and $0.2$.
The upper-left and upper-right panels show the numerical results 
for the type 1 initial condition of with $g=0.1$ and $0.2$, respectively,
while that of the lower panels for the type 2 initial condition
with the same $g$.}
\label{Fig:number_long}
\end{figure}

To quantify the above observation, we analyze the effective particle-numbers,
since one of the proposed mechanisms of thermalization
is the decay of the Yang-Mills field to particles.
We define an effective particle-number $N_\nu$
as a function of the longitudinal momentum $\nu$ integrated over the transverse momenta as, 
\begin{align}
N_\nu 
\equiv&\sum_{\bm{k}_\perp,\lambda,a}\left\{
\average{
\left(\hat{a}^{a\lambda\dagger}_{\bk}
-\langle\hat{a}^{a\lambda\dagger}_{\bk}\rangle\right)
\left(\hat{a}^{a\lambda}_{\bk}
-\langle\hat{a}^{a\lambda}_{\bk}\rangle\right)
-\frac{1}{2}}\right\}
\nonumber\\
=&\sum_{\bm{k}_\perp,\lambda,a}\left\{
\average{\hat{a}^{a\lambda\dagger}_{\bk}\hat{a}^{a\lambda}_{\bk}}
-\average{\hat{a}^{a\lambda\dagger}_{\bk}}
 \average{\hat{a}^{a\lambda}_{\bk}}
-\frac{1}{2}\right\}\ .
\label{eq:eff-dis-number-long}
\end{align}
Here we regard the creation and annihilation operators
subtracted with their expectation values
as those operators for the particles
under the background classical field.
This will be a reasonable description,
since the particle number is counted to be zero for a coherent state,
where the HW entropy takes the minimum value.
It should be noted that the last term of $-1/2$ comes from
the semiclassical treatment and the uncertainty relation.
In a semiclassical treatment, we cannot distinguish the order of the operators
and the expectation value of the symmetrized operator (Weyl product)
is observed.
For example, the expectation value of the the number operator
$\langle\hat{a}^\dagger\hat{a}\rangle$ is calculated as
$\langle(\hat{a}^\dagger\hat{a}+\hat{a}\hat{a}^\dagger)/2\rangle
=\langle(\omega^2\hat{x}^2+\hat{p}^2)/(2\omega)\rangle$
for a harmonic oscillator,
and its minimum value is $1/2$ as long as the distribution respects
the uncertainty principle.
Thus the nonzero $N_\nu$ signals the entropy production,
when we start from a coherent state initial condition.

In Fig.~\ref{Fig:number_long}, we show the effective particle-number
as a function of the longitudinal momentum $\tilde{\nu}$ at several values of $\tau$.
At the initial state which is set up to be a coherent state, we find $N_\nu=0$
(not shown in the figure) as expected.
In the earlier stage with $Q_s\tau=10$ (red curves), particles with the far lower longitudinal 
momenta dominate over those in other momentum regions.
At later times, particles with higher longitudinal momenta start to increase.
The early entropy production at the lower longitudinal momenta shown before thus naturally understood 
to be associated with the low momentum particle production.

Comparing Fig.~\ref{Fig:entropy_long} and Fig.~\ref{Fig:number_long}, 
we find that the particle creation is associated with the HW entropy
production, and there are two distinct stages in the evolution of fields.
In the first stage, the particle number in the low longitudinal momentum region grows and 
the HW entropy from the low longitudinal momentum modes
increases rapidly. 
In the second stage, the effective particle-number at higher longitudinal momenta grows and the HW 
entropy of higher longitudinal momentum modes increases slowly.

\section{Summary}\label{Sec:Summary}
\ \\
We have investigated the possible  thermalization process of the highly occupied and weakly coupled 
Yang-Mills fields in the expanding geometry through a computation of the entropy,
 as given by the Husimi-Wehrl (HW) entropy, (an)isotropization of the pressure and the particle 
production within the semiclassical approximation:
The time evolution of the system was obtained by solving the equation of motion of 
the Wigner function with use of the test particle method; 
the Husimi function is obtained 
by smearing the evaluated Wigner function in the phase space.
The initial condition of the simulation was constructed so as to mimic the glasma initial 
condition~\cite{MV1994_1,MV1994_2,MV1994_3,LMW1995_1,LMW1995_2}, where the macroscopic color electric 
and color magnetic fields are boost invariant and parallel to the collisional axis.
As such, we have considered two types of initial condition whose momentum
distributions are different from each other.

To obtain the HW entropy $\shw$ defined in terms of the Husimi function, it was first calculated by
two different test particle methods, the one is called "single test particle method (sTP)" and 
the other is called "parallel test particle method (pTP)". The resultant values thus obtained are denoted by 
$\shwp$  and $\shws$, respectively, and are shown to satisfy the inequalities $\shws<\shw<\shwp$.
We have shown that the average value of them, $\shwa$, turns out to give an excellent estimate 
of $\shw$ with numerical errors of $\mathcal{O}(\NTP^{-2})$ with $\NTP$ being the number of test 
particles, $\shwa = \shw + \mathcal{O}(\NTP^{-2})$,
while $\shws$ and $\shwp$ have numerical errors of $\mathcal{O}(\NTP^{-1})$
, $S^{{\rm sTP}({\rm pTP})}_{\rm HW} = \shw + \mathcal{O}(\NTP^{-1})$.
Nevertheless, to circumvent the computational difficulty in calculating the multiple integration in 
the HW entropy. 
we have taken the product ansatz for the Wigner function. 

Before proceeding with the study of the Yang-Mills field in the expanding geometry,
the above computational method was checked by applying it to the massless $\phi^4$ scalar theory 
in Minkowski space-time. 
The numerical results in the scalar theory have indicated that $\shws<\shw<\shwp$ and 
$\shwa \sim \shw$ nicely hold. 
It has been found that $\shw$ increases rapidly, then the growth rate becomes moderate, and finally 
stops increasing and keeps almost a constant value.
The saturation of $\shw$ indicates the achievement of equilibration of the system.

We have investigated the dynamical production of the HW entropy in the semiclassical evolution of the 
Yang-Mills field in the expanding geometry at $g=0.1$ and $0.2$. 
We have also shown the semiclassical evolution of the transverse and longitudinal pressures, 
$P_\perp$ and $P_\eta$.
Up to $Q_s\tau<(2-3)$, $P_\perp$ and $P_\eta$ 
have been found to approach each other, and there is no $g$
dependence at this time region.
In the later stage with $Q_s\tau>(2-3)$, 
$P_\perp$ and $P_\eta$ have been found
to approach some constant values slowly showing oscillatory behavior.
The amplitude of the oscillation becomes smaller 
at large $Q_s \tau$. 
The longitudinal pressure relative 
to the transverse pressure, $P_\eta/P_\perp$, 
has been found to come slightly closer as $g$ increases, 
but a pressure isotropization was not achieved, which is not in contradiction with a previous 
work~\cite{PressureEG2013} where a much larger coupling constant was used. 
After the earliest stage with $Q_s\tau<(2-3)$, where the HW entropy in the expanding geometry hardly 
increases, it grows rapidly in the following time range, $Q_s\tau<(20-30)$, and 
then increases more slowly in the later stage with $Q_s\tau>(20-30)$. 
For 
both types of initial conditions, the growth rate of the HW entropy at $g=0.2$ is larger 
than that at $g=0.1$. 
The slow HW entropy production stage does not readily mean that the system is near equilibrium since 
the large anisotropy of the pressure still remains in our simulations.
Such a slow production far from equilibrium is expected to be caused by the longitudinal expansion 
effect. 

We have defined the effective particle-number so that it represents the particle number created due to 
the deviations from the coherent state, and have compared its time evolution and the time evolution of 
the HW entropy, in piecewise of the longitudinal momentum mode. 
We have found that the effective particle-number and the HW entropy productions are 
associated with each other, and there are two distinct time stages in the time evolution of the 
Yang-Mills fields:
In the first stage, the particle number distribution in the low longitudinal momentum region grows 
and the HW entropy of the longitudinal low momentum modes increases rapidly, while in the second 
stage, the particle number distribution at higher longitudinal momentum grows 
and the HW entropy of corresponding modes increases slowly.

Since our choice of the initial conditions only mimic the glasma state,
we should directly take the McLerran-Venugopalan model in order to make the model more realistic~\cite{MVini}.
It is also a rather urgent subject to perform calculations at larger coupling constant such as
$g=0.5$ that is used in the previous calculation~\cite{PressureEG2013}, thereby clarifying the 
coupling dependence of the way of the thermalization process.

\section*{Acknowledgments}
This work is supported in part by Grants-in-Aid for Scientific Research from
Japan Society for the Promotion of Science (JSPS) (Nos. 19K03872, 19H01898, 19H05151, and 21H00121)
and by the Yukawa International Program for Quark–Hadron Sciences (YIPQS).

\appendix
\section{Second quantization of free gauge field in expanding geometry}\label{App:sc_exp}
In this Appendix, we present the second-quantized formulation of a free gauge field in the 
$\tau$-$\eta$ coordinate, which is found convenient to analyze fluctuations in the expanding 
glasma~\cite{Berges2014_2,flu_EG,Ini_flu}. 
This appendix refers to \cite{Berges2014_2}. 

First, we begin by showing the way of the second quantization in the continuum limit. 
In this paragraph only, all quantities are not normalized by lattice spacings. 
The equation of motion for the free gauge field, $[D,F]=0$, describing the $\tau$ evolution reads
\begin{align}
&\partial_1 \partial_\tau A_1(x)+\partial_2 \partial_\tau A_2(x)+\frac{1}{\tau^2}\partial_\eta\partial_\tau  A_\eta(x)=0\ ,\label{Eq:eom1}\\
&\left( \partial^2_{\tau}+\frac{1}{\tau}\partial_{\tau}- \partial^2_1 - \partial^2_2-\frac {\partial^2_\eta} {\tau^2} \right) A_i(x)
 +\partial_i\left( \partial_1A_1(x) + \partial_2A_2(x) + \frac{1}{\tau^2}\partial_\eta A_\eta(X) \right)=0\ ,\label{Eq:eom2}\\
&\left( \partial^2_{\tau}-\frac{1}{\tau}\partial_{\tau}-\partial^2_1 - \partial^2_2-\frac {\partial^2_\eta} {\tau^2} \right) A_\eta(x)
 +\partial_\eta\left( \partial_1A_1(x) + \partial_2A_2(x) + \frac{1}{\tau^2}\partial_\eta A_\eta(X) \right)=0 .\label{Eq:eom3}
\end{align}
For the Fourier modes with finite transverse momentum mode, 
the general solution satisfying the Coulomb type gauge condition, 
$\left( \partial_1 A_1 + \partial_2 A_2 + \tau^{-2}\partial_\eta A_\eta \right)\Bigl|_{\tau=\tau_0}=0$, 
is expressed in terms of the Hankel function,
\begin{align}
&A_{\mu}=\sum_{\lambda=1,2} \int \frac{d^3\bk}{(2\pi)^3} 
\left( \xi^{\lambda,(1)}_{\bk} \mathcal{A}^{(-)\lambda}_{\mu,\bk}(\tau)e^{i(\bk_\perp\cdot\bx_\perp + \nu\eta)} +
       \xi^{\lambda,(2)}_{\bk} \mathcal{A}^{(+)\lambda}_{\mu,\bk}(\tau)e^{i(\bk_\perp\cdot\bx_\perp + \nu\eta)} \right)\ ,\\
&\mathcal{A}^{(^-/_+)1}_{\mu,\bk}(\tau)=\frac{i}{k_\perp}(0, k_2, -k_1, 0)H^{(^1/_2)}_{i\nu}(k_\perp\tau)\ ,\\
&\mathcal{A}^{(^-/_+)2}_{\mu,\bk}(\tau)=-\frac{\nu}{k_\perp}
\left(0, k_1\alpha^{(^1/_2)}_{i\nu}(k_\perp\tau), k_2\alpha^{(^1/_2)}_{i\nu}(k_\perp\tau), 
-\frac{1}{\nu}\beta^{(^1/_2)}_{i\nu}(k_\perp\tau) \right)\ ,\\
&\alpha^{(^1/_2)}_{i\nu}(k_\perp\tau)=\int^{k_\perp\tau}_{k_\perp\tau_0} dz 
 \frac{1}{z}H^{(^1/_2)}_{i\nu}(z)-\frac{ k_\perp\tau_0   }{\nu^2+(k_\perp\tau_0)^2}\dot{H}^{(^1/_2)}_{i\nu}(k_\perp\tau_0)\ ,\\
&\beta ^{(^1/_2)}_{i\nu}(k_\perp\tau)=\int^{k_\perp\tau}_{k_\perp\tau_0} dz
          z H^{(^1/_2)}_{i\nu}(z)-\frac{(k_\perp\tau_0)^3}{\nu^2+(k_\perp\tau_0)^2}\dot{H}^{(^1/_2)}_{i\nu}(k_\perp\tau_0)\ ,
\end{align}
where $k_\perp=\sqrt{k^2_1+k^2_2}$ is a transverse momentum 
and $\xi^{\lambda,i}_{\bk}$ is a given constant. 
The solution, $\mathcal{A}^{(i)\lambda}_{\mu,\bk}(\tau)e^{i(\bk_\perp\cdot\bx_\perp + \nu\eta)}$, is 
orthogonal to other solution having different indexes 
$(\lambda', i', \bk')\neq(\lambda, i, \bk)$, 
\begin{align}
\left( \mathcal{A}^{(i)\lambda}_{\mu,\bk}(\tau)e^{i\bk\cdot\bx}, \mathcal{A}^{(i')\lambda}_{\mu',\bk'}(\tau)e^{i\bk'\cdot\bx}\right)=0\ ,
\end{align}
with respect to a scalar product defined as 
\begin{align}
\left( f_\mu(x),g_\mu(x)\right)=-i\int d^3 \bx \tau g^{\mu\nu} \left( f^*_\mu(x)\partial_\tau g_\nu(x) - g_\nu(x)\partial_\tau f^*_\mu(x) \right)\ .
\end{align}
If $(\lambda', i', \bk')=(\lambda, i, \bk)$, the scalar product is not vanished and is invariant under 
the $\tau$ evolution according to Eq.~\refeq{Eq:eom1}-\refeq{Eq:eom3}. 
Since the second kind of Hankel function asymptotically behaves as the positive frequency mode,
\begin{align}
H^{(i)}_{i\nu}(k_\perp\tau)\to\sqrt{\frac{2}{\pi k_\perp\tau}}
\exp{\left[-i\left(k_\perp\tau-\frac{\pi}{4}\right)-\frac{\pi \nu}{2}\right]}\ \ \ (k_\perp\tau\rightarrow\infty)\ ,
\end{align}
we can obtain the expression of the second-quantized gauge field as the linear combination of 
$\mathcal{A}^{(+)1}_\mu(\tau,\bk)$ and $\mathcal{A}^{(+)2}_\mu(\tau,\bk)$, 
\begin{align}
&A_i(x)=\sum_\lambda\int\frac{d^3\bk}{(2\pi)^3}
\left(  \hca^\lambda_{\bk}\mathcal{A}^{\lambda}_{\mu,\bk}(\tau) e^{i(\bk_\perp\cdot\bx_\perp+\nu\eta)} + {\rm h.c.}\right)\ ,\\
&\mathcal{A}^{\lambda}_{\mu,\bk}(\tau) = \sqrt{\frac{\pi}{4}}e^{\frac{\pi |\nu|}{2}} \mathcal{A}^{(+)\lambda}_{\mu,\bk}(\tau)\ ,\label{Eq:nom}\\
&\left[ \hca^\lambda_{\bk}, \hca^{\lambda'\dagger}_{\bk'} \right] = (2\pi)^3\delta_{\lambda,\lambda'}\delta(\bk-\bk')\ .
\end{align}
Here we determine the normalization constant in front of $\mathcal{A}^{(+)\lambda}_{\mu,\bk}(\tau)$ 
in Eq.~\refeq{Eq:nom} so as to satisfy the ortho-normal condition, 
\begin{align}
&\left( \sqrt{\frac{\pi}{4}}e^{\frac{\pi |\nu|}{2}} \mathcal{A}^{(+)\lambda}_{\mu,\bk}(\tau)e^{i(\bk_\perp\cdot\bx_\perp+\nu\eta)}, 
\sqrt{\frac{\pi}{4}}e^{\frac{\pi |\nu'|}{2}} \mathcal{A}^{(+)\lambda'}_{\mu,\bk'_\perp,\nu'}(\tau)e^{i(\bk'_\perp\cdot\bx_\perp+\nu'\eta)}\right)
\nonumber\\
&=(2\pi)^3\delta_{\lambda,\lambda'}\delta(\bk_\perp-\bk'_\perp)\delta(\nu-\nu')\ , \label{Eq:nor}
\end{align}
The procedure of the second quantization using such a scalar product is standard in Quantum field theory in curved space-time.
In our analysis, we only treat the finite transverse momentum modes since the contribution of the $0$ transverse mode decreases 
as the transverse size of the system becomes large. 

Next, we show the second quantization of the free gauge field on the space lattice with the 
continuous time.
The equation of motion reads
\begin{align}
&\partial^{\rm B}_1 \partial_\tau A_1(x)+\partial^{\rm B}_2 \partial_\tau A_2(x)
+\frac{1}{a^2_\eta\tau^2}\partial^{\rm B}_\eta\partial_\tau  A_\eta(x)=0\ ,\label{Eq:eom1_lat}\\
&\left( \partial^2_{\tau}+\frac{1}{\tau}\partial_{\tau} - \partial^{\rm B}_1\partial^{\rm F}_1 - \partial^{\rm B}_2\partial^{\rm F}_2
-\frac {\partial^{\rm B}_\eta\partial^{\rm F}_\eta} {a^2_\eta\tau^2} \right) A_i(x)\nonumber\\
&\ \ \ \ \ \ \ \ \ \ +\partial_i\left( \partial^{\rm B}_1A_1(x) + \partial^{\rm B}_2A_2(x) + \frac{1}{a^2_\eta\tau^2}\partial^{\rm B}_\eta A_\eta(X) \right)=0\ ,\label{Eq:eom2_lat}\\
&\left( \partial^2_{\tau}-\frac{1}{\tau}\partial_{\tau}- \partial^{\rm B}_1\partial^{\rm F}_1 - \partial^{\rm B}_2\partial^{\rm F}_2
-\frac {\partial^{\rm B}_\eta\partial^{\rm F}_\eta} {a^2_\eta\tau^2} \right) A_\eta(x)\nonumber\\
&\ \ \ \ \ \ \ \ \ \ +\partial_i\left( \partial^{\rm B}_1A_1(x) + \partial^{\rm B}_2A_2(x) + \frac{1}{a^2_\eta\tau^2}\partial^{\rm B}_\eta A_\eta(X) \right)=0\ ,\label{Eq:eom3_lat}
\end{align}
where $\partial^{\rm B}_i$ denotes a backward difference operator in the $i$-direction. 
In much the same way as the continuum case, 
we can get the expression of the second-quantized gauge field as
\begin{align}
&A_i(x)=\frac{1}{\sqrt{L^2_\perp L_\eta}}\sum_{\lambda,\bk} 
\left( \hca^\lambda_{\bk}\tilde{\mathcal{A}}^{\lambda}_{\mu,\bk}(\tau) e^{i(\bk_\perp\cdot\bx_\perp+\nu\eta)} + {\rm h.c.}\right)\ ,\\
&\left[\hca^\lambda_{\bk},\hca^{\lambda'\dagger}_{\bk} \right] = \delta_{\lambda,\lambda'}\delta_{\bk,\bk'}\ ,\\
&\tilde{\mathcal{A}}^{1}_{\mu,\bk}(\tau)=\frac{i}{\omega_{\bm{k}_\perp}}                    \sqrt{\frac{\pi}{4\al}}e^{\frac{\pi |\tilde{\nu}|}{2\al}}
(0, \tilde{k}_2, \tilde{k}_1, 0)H^{(2)}_{i|\tilde{\nu}|/\al}(\omega_{\bm{k}_\perp}\tau)\ ,\\
&\tilde{\mathcal{A}}^{2}_{\mu,\bk}(\tau)=-\frac{\tilde{\nu}^*}{\al \omega_{\bm{k}_\perp}}\sqrt{\frac{\pi}{4\al}}e^{\frac{\pi |\tilde{\nu}|}{2\al}} 
\left(0, \tilde{k}_1\alpha^{(2)}_{i|\tilde{\nu}|/\al}(\omega_{\bm{k}_\perp}\tau), \tilde{k}_2\alpha^{(2)}_{i|\tilde{\nu}|/\al}(\omega_{\bm{k}_\perp}\tau), 
-\frac{\al^2}{\tilde{\nu}^*}\beta^{(2)}_{i|\tilde{\nu}|/\al}(\omega_{\bm{k}_\perp}\tau) \right)\ .
\end{align}
By utilizing this expression, 
the electric field and the free part of the magnetic field are also written in terms of the annihilation and creation operators,
\begin{align}
&\hat{E}^i(x)  =\frac{1}{\sqrt{L^2_\perp L_\eta}}\sum_{\lambda,\bk} 
                \left( \hca^\lambda_{\bk}\mathcal{E}^{\lambda,i}_{\bk}(\tau) e^{i(\bk_\perp\cdot\bx_\perp+\nu\eta)} + {\rm h.c.}\right)\ ,\label{Eq:sq_e}\\
&\hat{B}^i_0(x)=\frac{1}{2}\epsilon^{ijk}\partial^{\rm F}_j A_k(x)
               =\frac{1}{\sqrt{L^2_\perp L_\eta}}\sum_{\lambda,\bk} 
                \left( \hca^\lambda_{\bk}\mathcal{B}^{\lambda,i}_{\bk}(\tau) e^{i(\bk_\perp\cdot\bx_\perp+\nu\eta)} + {\rm h.c.}\right)\ ,\label{Eq:sq_b}\\
&\mathcal{E}^{1,i}_{\bk}(\tau)=\dot{H}^{(2)}_{i|\tilde{\nu}|/\al}(\omega_{\bm{k}_\perp}\tau)\varepsilon^{1,i}_{\bk}\ ,\ \ 
 \mathcal{E}^{2,i}_{\bk}(\tau)=     H ^{(2)}_{i|\tilde{\nu}|/\al}(\omega_{\bm{k}_\perp}\tau)\varepsilon^{2,i}_{\bk}\ ,\\
&\mathcal{B}^{1,i}_{\bk}(\tau)=     H ^{(2)}_{i|\tilde{\nu}|/\al}(\omega_{\bm{k}_\perp}\tau)\varepsilon^{2,i*}_{\bk}\ ,\ \ 
 \mathcal{B}^{2,i}_{\bk}(\tau)=\dot{H}^{(2)}_{i|\tilde{\nu}|/\al}(\omega_{\bm{k}_\perp}\tau)\varepsilon^{1,i*}_{\bk}\ ,\\
&\varepsilon^{1,i}_{\bk} =i\al\tau\sqrt{\frac{\pi}{4\al}}e^{\frac{\pi |\tilde{\nu}|}{2\al}}(\tilde{k}^*_2, -\tilde{k}^*_1, 0)\ ,\ \ 
 \varepsilon^{2,i}_{\bk} =- \frac{\tilde{\nu}^*}{\omega_{\bm{k}_\perp}}\sqrt{\frac{\pi}{4\al}}e^{\frac{\pi |\tilde{\nu}|}{2a_\eta}}
 \left(\tilde{k}_1, \tilde{k}_2, -\frac{\omega^2_{\bm{k}_\perp}}{\tilde{\nu}^*} \right)\ .
\end{align}
Solving Eq.~\refeq{Eq:sq_e} and Eq.~\refeq{Eq:sq_b} with respect to $\hca^\lambda_{\bk}$, 
we can write $\hca^\lambda_{\bk}$ as the linear combination of the Fourier modes, $(\hat{E}_{\bk}, \hat{B}_{\bk})$,
\begin{align}
\hca^1_{\bk}&=\frac{i}{\al \omega_{\bk_\perp} \tau}\nonumber\\
&\times \left(  
 \frac{\omega^2_{\bk_\perp}}{\omega^2_{\bm{k}_\perp}+|\tilde{\nu}/(\tau\al)|^2} 
 \frac{\dot{H}^{(2)*}_{i|\tilde{\nu}|/\al}(\omega_{\bm{k}_\perp} \tau)}{H^{(2)*}_{i|\tilde{\nu}|/\al}(\omega_{\bm{k}_\perp} \tau)}
 \left[\mathcal{B}^{1*}_{\bk} \cdot  B_{0,\bk}(\tau)\right] 
+\frac{H^{(2)*}_{i|\tilde{\nu}|/\al}(\omega_{\bm{k}_\perp} \tau)}{\dot{H}^{(2)*}_{i|\tilde{\nu}|/\al}(\omega_{\bm{k}_\perp} \tau)} 
 \left[\mathcal{E}^{1,*}_{\bk} \cdot E_{\bk}(\tau) \right] \right)\ ,\label{Eq:sq_a1}\\
\hca^2_{\bk}&=\frac{i}{\al \omega_{\bm{k}_\perp} \tau}\nonumber\\
&\times \left(  
 \frac{\omega^2_{\bk_\perp}}{\omega^2_{\bm{k}_\perp}+|\tilde{\nu}/(\tau\al)|^2} 
 \frac{\dot{H}^{(2)*}_{i|\tilde{\nu}|/\al}(\omega_{\bm{k}_\perp} \tau)}{H^{(2)*}_{i|\tilde{\nu}|/\al}(\omega_{\bm{k}_\perp} \tau)}
 \left[\mathcal{E}^{2*}_{\bk} \cdot  E_{\bk}(\tau)\right] 
+\frac{H^{(2)*}_{i|\tilde{\nu}|/\al}(\omega_{\bm{k}_\perp} \tau)}{\dot{H}^{(2)*}_{i|\tilde{\nu}|/\al}(\omega_{\bm{k}_\perp} \tau)} 
 \left[\mathcal{B}^{2*}_{\bk} \cdot B_{0,\bk}(\tau) \right] \right)\ .\label{Eq:sq_a2}
\end{align}
In actual calculations, we use these relations to transform $(A,E)$ to $(\Phi,\Pi)$ 
through Eq.~\refeq{Eq:phidot2} and Eq.~\refeq{Eq:pidot2}.

\section{Divergence in pressure and energy density}\label{App:ini_div}
In this appendix, we show the remaining divergences in the pressure and energy density after 
subtracting the vacuum contribution, and discuss how to subtract them. 
For convenience, we use the following notation $()_{\rm ini\ mac}$, 
which means the contribution of macroscopic field part of the initial Wigner function to a given observable at the initial time $\tau_0$. 
For example, in the case of the Fourier transforms of the color electric and color magnetic fields, 
$(\bm{E}^a_{\bk})_{\rm ini\ mac}$ and $(\bm{B}^a_{\bk})_{\rm ini\ mac}$ have been already given in Eq.~\refeq{mac:e1}-\refeq{mac:b3}. 
Thus, the  the macroscopic field contribution to the initial pressure are given by
\begin{align}
 (P_\eta )_{\rm ini\ mac}\sim
-(P_\perp)_{\rm ini\ mac}&=    -\frac{1}{V} \sum_x g_{ii} \left(\hat{T}^{ii}(x)\right)_{\rm ini\ mac} \nonumber\\
                         &=    -\frac{a_\eta\tau}{2} \sum_{a,\bm{x}_\perp,\eta} 
                                \left[ \left( \hat{E}^{a\eta}(x)^2 \right)_{\rm ini\ mac} + \left( \hat{B}^{a\eta}(x)^2 \right)_{\rm ini\ mac} \right]\nonumber\\
                         &\sim -\frac{\Delta}{8L^2_\perp \alpha_s}\sum_{\bk_\perp} \omega^2_{\bm{k}_\perp} |f_l(\bm{k}_\perp)|^2\ .
\end{align}
                    
\begin{figure}[hbtp]
\begin{minipage}{0.5\hsize}
\centering\includegraphics[width=80mm,bb=0 0 360 252]{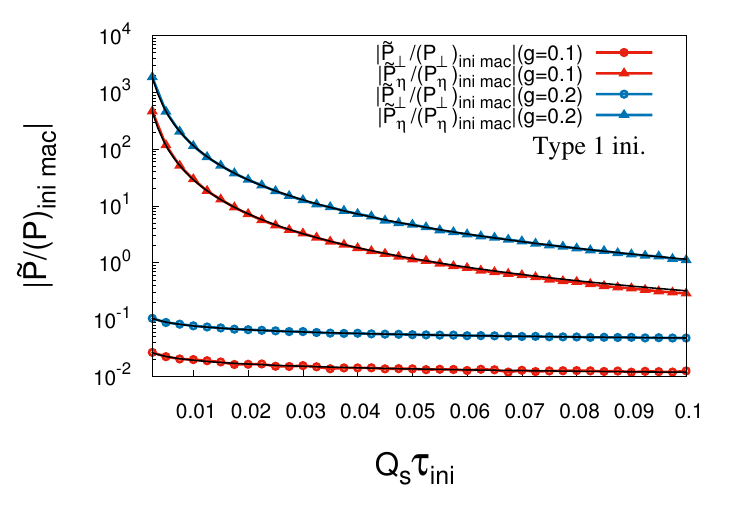}\\
\end{minipage}
\begin{minipage}{0.5\hsize}
\centering\includegraphics[width=80mm,bb=0 0 360 252]{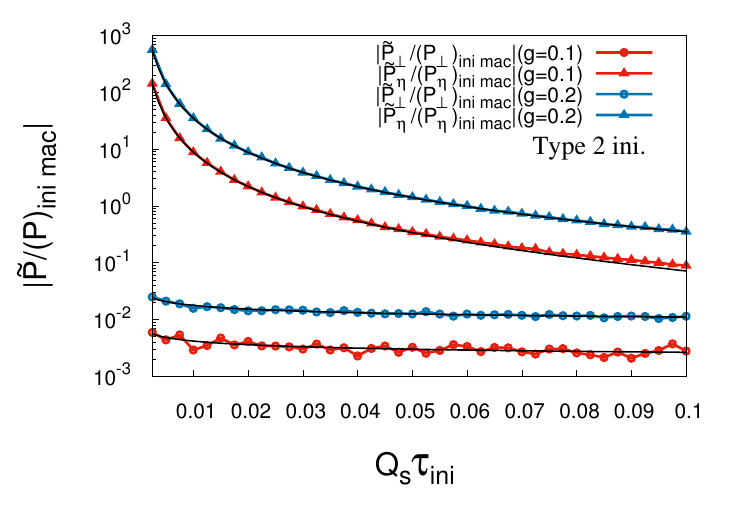}\\
\end{minipage}
\caption{
The $Q_s\tau_0$ dependence of the initial pressure after subtracting the vacuum contribution 
and the macroscopic field contribution, 
$\tilde{P}_i = \frac{1}{V}\sum_x g_{ii} \average{\hat{T}^{ii}(x)}_{\rm mat+flu} - \frac{1}{V}\sum_x g_{ii} \average{\hat{T}^{ii}(x)}_{\rm vac} 
            - (P_i)_{\rm ini\ mac}$, normalized by $(P_i)_{\rm ini\ mac}$.
These calculations are performed at at $g=0.1$ and $0.2$ for two types of initial condition. 
The red (blue) 
line with circles shows the transverse pressure at $g=0.1$ $(0.2)$. 
The red (blue) 
line with triangular points shows longitudinal one at $g=0.1$ $(0.2)$. 
The result in the left (right) panel is calculated with the type $1$ ($2$) initial condition. 
}\label{Fig:ini}
\end{figure}
In Fig.~\ref{Fig:ini}, we show the $Q_s\tau_0$ dependence of the initial pressure after subtracting 
the vacuum contribution and the macroscopic field contribution, 
\begin{align}
\tilde{P}_i = \frac{1}{V}\sum_x g_{ii} \average{\hat{T}^{ii}(x)}_{\rm mat+flu} - \frac{1}{V}\sum_x g_{ii} \average{\hat{T}^{ii}(x)}_{\rm vac} 
            - (P_i)_{\rm ini\ mac}\ ,
\end{align}
normalized by $(P_i)_{\rm ini\ mac}$ at $g=0.1$ and $0.2$ for two types of initial condition. 
We find that $\left|\tilde{P}_\eta/(P_\eta)_{\rm ini\ mac}\right|$ diverges as $(Q_s\tau_0)^{-2}$ 
since the $Q_s\tau_0$ dependence of $\left|\tilde{P}_\eta/(P_\eta)_{\rm ini\ mac}\right|$ is reproduced well by the 
fitting function $f_{\rm pow}(Q_s\tau_0)=A_{\rm pow} / (Q_s\tau_0)^2$. 
We also find that $\left|\tilde{P}_\perp/(P_\perp)_{\rm ini\ mac}\right|$ diverges as 
$\ln^2{Q_s\tau_0}$ since $\left|\tilde{P}_\perp/(P_\perp)_{\rm ini\ mac}\right|$ data lie on the 
fitting curve $f_{\rm log}(Q_s\tau_0)=B_{\rm log} \ln^2{(Q_s\tau_0)} + C_{\rm log}$.

Here, we briefly explain how to remove the remaining divergence at any $\tau$ based on the method 
presented in Ref.~\cite{PressureEG2013}. 
As seen above, the remaining divergence in $P_\perp$ at $\tau=\tau_0$ is a logarithmic function of 
$\tau_0$ and is much smaller than the initial macroscopic field contribution,
\begin{align}
B_{\rm log} \ln^2{(Q_s\tau_0)} \gg |(P_\perp)_{\rm ini\ mac}|\ .
\end{align}
Thus, we assume to neglect the remaining divergence in $P_\perp$ at any $\tau$. 
Then we can define the subtracted energy density and pressure as
\begin{align}
&\varepsilon \equiv \frac{1}{V}\sum_x g_{\tau\tau} \average{\hat{T}^{\tau\tau}(x)}_{\rm mac+fluc} - \frac{1}{V}\sum_x g_{\tau\tau} \average{\hat{T}^{\tau\tau}(x)}_{\rm vac} -                \alpha_{\rm div}(t)\ ,\\
&P_i         \equiv \frac{1}{V}\sum_x g_{ii      } \average{\hat{T}^{ii      }(x)}_{\rm mac+fluc} - \frac{1}{V}\sum_x g_{ii      } \average{\hat{T}^{ii      }(x)}_{\rm vac} - \delta_{i,\eta}\alpha_{\rm div}(t)\ ,
\end{align}
where $\alpha_{\rm div}(\tau)$ represents the divergence that should be removed. 
The energy density also has the remaining divergence because of the relation between the energy density
and pressure, $\varepsilon=P_1+P_2+P_\eta$. 
By using the conservation law in the boost invariant longitudinal (Bjorken) expanding geometry,
$\partial \varepsilon/\partial \tau=-\varepsilon+P_\eta/\tau$, 
we obtain the evolution equation for $\alpha$, 
\begin{align}
\frac{d \alpha_{\rm div}}{d \tau}=-\frac{2\alpha_{\rm div}}{\tau}\ .
\end{align}
By solving the differential equation, we find $\alpha_{\rm div}(\tau) = \alpha /\tau^2$, where $\alpha$ is a constant.
As seen above, we can obtain $\alpha(\tau_0)$ as $\alpha(\tau_0)=A_{\rm pow}/(Q_s\tau_0)^2$ by the fit.
Thus, we use $\alpha=A_{\rm pow}$ in actual calculations.

\section{Evaluation by test particle methods}\label{App:ne_tpm}
An integral $I$ consisting of a function $H(\Gamma)$
that is evaluated by test particle methods
is expressed as
\begin{eqnarray}
I = \int  \dd \Gamma {\cal F}(H(\Gamma)),
\end{eqnarray}
where $\Gamma$ denotes the phase space point under consideration.
In general, $H(\Gamma)$ evaluated with $i$-th set of test particles,
which is represented as $H_i(\Gamma)$,
has numerical errors $dH_i(\Gamma)$ depending on the phase space point $\Gamma$
as
\begin{eqnarray}
H_i(\Gamma) = H(\Gamma)+dH_i(\Gamma).
\end{eqnarray}
Note that each $H(\Gamma)$ that enters the integral $I$ can be evaluated
with different and independent sets of test particles.
(When all the test particle sets are identical, it is called the single test particle method.)
Then, under the condition $\frac{dH}{H}\ll 1$,
the integral $I$ can be expanded as a series of $dH_i(\Gamma)$,
\begin{eqnarray}
 I=\int  \dd \Gamma {\cal F}(H)
  +
  \sum_{i1,i2,..,iN} \int \dd\Gamma {\cal F'}^{1,2,..,N}_{i1,i2,..,iN}dH_1^{i1}dH_2^{i2}..dH_N^{iN},
\end{eqnarray}
where ${\cal F'}$ denotes coefficients in the expansion,
and $\Gamma$ in the functions are omitted.
We here consider the situation where
the integrals of the odd-order terms of $dH$ disappear
due to numerical error cancellation
as $\int \dd\Gamma dH_i^{2n+1}(\Gamma){\cal G}(\Gamma) = 0$,
which would be justified when positive and negative contributions 
of $dH_i(\Gamma)$ equally enter in the integration as 
$\int {\rm d} \Gamma dH_i(\Gamma) = 0$ and ${\cal G}$ is smooth enough.
In such a case, only even-order terms contribute to $I$'s numerical
errors as
\begin{eqnarray}
 I=\int  \dd \Gamma {\cal F}(H)
  +
  \sum_{i1,i2,..,iN\in{\rm even}} \int \dd\Gamma {\cal F'}^{1,2,..,N}_{i1,i2,..,iN}dH_1^{i1}dH_2^{i2}..dH_N^{iN},
\end{eqnarray}
and the number of terms is greatly reduced.

Let us proceed with the evaluation of a HW entropy based on test particle methods.
A HW-entropy $S$ can be expressed as
\begin{eqnarray}
S=-\int d\Gamma H(\Gamma)\ln H(\Gamma),
\end{eqnarray}
with a Husimi function $H(\Gamma)$.
With the parallel test particle (pTP) method in mind,
$S$ can be written as
\begin{eqnarray}
 S
 &=&
 -\int d\Gamma H_1(\Gamma)\ln H_2(\Gamma)\nonumber\\
 &=&
 -\int d\Gamma (H(\Gamma)+dH_{1}(\Gamma))\ln (H(\Gamma)+dH_{2}(\Gamma))\ ,
\end{eqnarray}
which is further expanded as
\begin{eqnarray}
 S=-\int \dd\Gamma H(\Gamma)\ln H(\Gamma)
  -\int \dd\Gamma dH_{1}(\Gamma)\ln H(\Gamma)
  -\int \dd\Gamma dH_{2}(\Gamma)
\nonumber\\
  -\int \dd\Gamma \frac{dH_{1}(\Gamma)dH_{2}(\Gamma)}{H(\Gamma)}
  +\int \dd\Gamma \frac{dH_{2}^2(\Gamma)}{2H(\Gamma)}
  +\int \dd\Gamma \frac{dH_{1}(\Gamma)dH_{2}^2(\Gamma)}{2H(\Gamma)}+\cdots\ ,
\end{eqnarray}
when $\frac{dH}{H}\ll 1$.
In the case where $H(\Gamma)$ is a Husimi function,
we have confirmed the equality $\int \dd \Gamma H(\Gamma) = 1$ numerically holds in a good accuracy,
which indicates that positive and negative contributions of $dH_i(\Gamma)$ equally enter in the integration, and
the integrals of the odd-order terms of $dH$ is expected to disappear
due to numerical error cancellation
as $\int {\rm d}\Gamma dH^{2n+1}(\Gamma){\cal G}(\Gamma) = 0$,
since ${\cal G}(\Gamma)$ is a function of $H(\Gamma)$, which is a smooth Gaussian-smeared function.
Such error cancellation has been numerically confirmed 
at least in the calculations presented in this paper,
and we can leave only even-order terms of $dH$.

For the single test particle (sTP) method,
the errors in and outside the logarithmic function are identical
($dH_1=dH_2$) and $S$ leads to the form,
\begin{eqnarray}
 S^{\rm sTP}=-\int \dd\Gamma H(\Gamma)\ln H(\Gamma)
  -\int \dd\Gamma \frac{dH_{2}^2(\Gamma)}{2H(\Gamma)}\ .
\end{eqnarray}
For the parallel test particle (pTP) method,
$dH_1$ and $dH_2$ are independent, and
\begin{eqnarray}
 S^{\rm pTP}=-\int \dd\Gamma H(\Gamma)\ln H(\Gamma)
  +\int \dd\Gamma \frac{dH_{2}^2(\Gamma)}{2H(\Gamma)}
\end{eqnarray}
holds.
We finally obtain the inequality,
\begin{eqnarray}
S^{\rm sTP}  < S < S^{\rm pTP}.
\end{eqnarray}
When the test particles used for evaluating $\ln H(\Gamma)$ are common in sTP and pTP methods,
the entropy evaluated with $\NTP\rightarrow\infty$ can be obtained as
\begin{eqnarray}
S = \frac{S^{\rm sTP}+S^{\rm pTP}}{2} + \mathcal{O}(\NTP^{-2}), 
\end{eqnarray}
where we make the reasonable assumption, $dH \propto 1/\sqrt{\NTP}$.

\section{Choice of the smearing parameter}\label{App:ChoiceOfSigma}
Here we show how the HW entropy defined in Eq.~\refeq{Eq:HWs}, in which the smear parameters are set 
to the eigenfrequencies, behaves for the Gibbs state and vacuum state of the free field.

We first discuss the HW entropy defined in Eq.~\refeq{Eq:HWs} for the Gibbs 
ensemble of the free field, $\rho^{\rm free}_{\rm Gibbs} \propto e^{-H_{\rm free}/T}$.
In this case, the total density matrix, $\rho^{\rm free}_{\rm Gibbs}$, can be written as the product 
of the Gibbs ensembles of one-dimensional harmonic oscillators,
$\rho^{\rm free}_{\rm Gibbs} = \prod_{\bk} \rho^{\rm h.o.}_{{\rm Gibbs},\bk} \propto \prod_{\bk} e^{-H^{\rm h.o.}_{\bk}/T}$.
Then the total Husimi function can also be written as the product of the Husimi functions for each 
degree of freedom,
\begin{align}
&\fh(\{\Phi,\Pi,\omega\}) \Big|_{\rho=\rho^{\rm free}_{\rm Gibbs}}
= \prod_{\bk} \fh(\Phi_{\bk},\Pi_{\bk},\omega_{\bk}) \Big|_{\rho=\rho^{\rm h.o.}_{{\rm Gibbs},\bk}}\ .
\end{align}
Therefore, the total HW entropy, $\shw|_{\rho=\rho^{\rm free}_{\rm Gibbs}}$, is given by the sum of 
the HW entropy for each degree of freedom, 
\begin{align}
\shw(\{\omega\}) \Big|_{\rho=\rho^{\rm free}_{\rm Gibbs}}
&=-\int \mathcal{D}\Gamma \fh(\{\Phi,\Pi,\omega\})\ln{\fh(\{\Phi,\Pi,\omega\})}
\Big|_{\rho=\rho^{\rm free}_{\rm Gibbs}}\nonumber\\
&=\sum_{\bk} -\int \frac{d\Phi_{\bk}d\Pi_{\bk}}{2\pi}\fh(\Phi_{\bk},\Pi_{\bk},\omega_{\bk})\ln{\fh(\Phi_{\bk},\Pi_{\bk},\omega_{\bk})}
\Big|_{\rho=\rho^{\rm h.o.}_{{\rm Gibbs},\bk}}\nonumber\\
&=\sum_{\bk} \shw(\omega_{\bk}) \Big|_{\rho=\rho^{\rm h.o.}_{{\rm Gibbs},\bk}}\ .
\end{align}
On the basis of the discussion of the HW entropy of an one-dimensional harmonic oscillator with the 
smearing parameter being set to its eigenfrequency, which is given in Sec.~$5$ in 
Ref.~\cite{Kyoto2009}, the HW entropy for each degree of freedom, 
$\shw(\omega_{\bk}) \Big|_{\rho=\rho^{\rm h.o.}_{{\rm Gibbs},\bk}}$, is found to be larger than the 
von-Neumann entropy $S_{\rm vN}\left(=-{\rm Tr}(\rho^{\rm h.o.}_{{\rm Gibbs},\bk}\ln\rho^{\rm h.o.}_{{\rm Gibbs},\bk})\right)$ 
obtained from the same density matrix, but to agree with it in the high-temperature limit, 
\begin{align}
                   \shw(\omega_{\bk}) \Big|_{\rho=\rho^{\rm h.o.}_{{\rm Gibbs},\bk}} 
                   &>                    S_{\rm vN} \Big|_{\rho=\rho^{\rm h.o.}_{{\rm Gibbs},\bk}}\ ,\\
\lim_{T \to \infty} \shw(\omega_{\bk}) \Big|_{\rho=\rho^{\rm h.o.}_{{\rm Gibbs},\bk}} 
                   &= \lim_{T \to \infty} S_{\rm vN} \Big|_{\rho=\rho^{\rm h.o.}_{{\rm Gibbs},\bk}}\ .
\end{align}
Accordingly. the same relationship holds for for the total HW entropy and the von-Neumann entropy 
given as a sum of harmonic oscillators,
\begin{align}
                   \shw(\{\omega\}) \Big|_{\rho=\rho^{\rm free}_{\bk}} &>                    S_{\rm vN} \Big|_{\rho=\rho^{\rm free}_{\bk}}\ ,\\
\lim_{T \to \infty} \shw(\{\omega\}) \Big|_{\rho=\rho^{\rm free}_{\bk}} &= \lim_{T \to \infty} S_{\rm vN} \Big|_{\rho=\rho^{\rm free}_{\bk}}\ ,
\end{align}
where $S_{\rm vN} \Big|_{\rho=\rho^{\rm free}_{\bk}} = \sum_{\bk} S_{\rm vN} \Big|_{\rho=\rho^{\rm h.o.}_{{\rm Gibbs},\bk}}$.
This shows that the HW entropy $\shw(\{\omega_{\bk}\})$ adopted in our study agrees with the
von-Neumann entropy in the high-temperature and weak-coupling limit.

Next, we show that the HW entropy defined by Eq.~\refeq{Eq:HWs} takes the minimum value 1 for the 
perturbative vacuum state as
\begin{align}
\shw(\{\omega\})\geq \shw(\{\omega\})\Big|_{|0\rangle\langle 0|}= 1.
\end{align}
To show the above inequality, we utilize the theorem given in Refs.~\cite{Lieb1978,Carlen1990} 
stating that, in general, the HW entropy for given conjugate variables $(\hat{x},\hat{p})$ and 
smearing parameter $\sigma$ takes the minimum value $1$ for the coherent state 
$|\alpha;\sigma\rangle$ defined as the eigenstate of the 
''annihilation operator'', $\hca=(\sigma\hat{x}+i\hat{p})/\sqrt{2\sigma}$:
\begin{align}
\shw(\sigma)\geq \shw(\sigma)\Big|_{|\alpha;\sigma\rangle\langle\alpha;\sigma|}= 1.
\end{align}
Then, one sees that the HW entropy with the smearing width $\{\omega_{\bk}\}$ automatically takes 
the minimum value 1 for the coherent states $|\{\alpha_{\bk}\};\{\omega_{\bk}\}\rangle$ defined 
by the annihilation operators $\{\hat{a}_{\bk}\}$ given by
(\ref{eq:field-annih-cano-rel}), as
\begin{align}
\shw(\{\omega\})\geq \shw(\{\omega\})\Big|_{|\{\alpha_{\bk}\};\{\omega_{\bk}\}\rangle\langle\{\alpha_{\bk}\};\{\omega_{\bk}\}|}= 1.
\end{align}
Such coherent states $|\{\alpha_{\bk}\};\{\omega_{\bk}\}\rangle$ include the perturbative vacuum 
state $|0\rangle$ since it is an eigenstate of $\hat{a}_{\bk}$ ($\hat{a}_{\bk}|0\rangle = 0$),
and then
\begin{align}
\shw(\{\omega\})\geq \shw(\{\omega\})\Big|_{|0\rangle\langle 0|}= 1
\end{align}
generally holds.

\end{document}